\documentclass[useAMS,usenatbib]{mn2e}
\usepackage{graphicx}

\newcommand{\gtrsim}{\mbox{{\raisebox{-0.4ex}{$\stackrel{>}{{\scriptstyle\sim}}$}}}}
\newcommand\arcdeg{\mbox{$^\circ$}}
\newcommand{\lesssim}{\mbox{{\raisebox{-0.4ex}{$\stackrel{<}{{\scriptstyle\sim}}$}}}}
\newcommand\aj{AJ}
\newcommand\apj{ApJ}
\newcommand\mnras{MNRAS}
\newcommand\apjl{ApJ}
\newcommand\bain{Bull.~Astron.~Inst.~Netherlands}
\newcommand\apjs{ApJS}
\newcommand\araa{ARA\&A}
\newcommand\apss{Ap\&SS}
\newcommand\aap{A\&A}
\newcommand\nat{Nature}

\title[The Distance to SS\,433/W50 and its Interaction with 
the ISM]{The Distance to SS\,433/W50 and its Interaction with 
the Interstellar Medium}
\author[Lockman, Blundell \& Goss]{Felix J.\
  Lockman$^{1}$\thanks{E-mail: jlockman@nrao.edu}, Katherine
  M.\ Blundell$^{2}$
 and W.\ M.\  Goss$^{3}$\\
$^{1}$National Radio Astronomy Observatory, P.O.\ Box 2, Green Bank,
  WV 24944, USA\\
$^{2}$Oxford University Astrophysics, Keble Road, Oxford, OX1 3RH\\
$^{3}$National Radio Astronomy Observatory, P.O.\ Box 0, Socorro, NM
  87801, USA}

\begin{document}

\date{to appear in MNRAS}

\pagerange{\pageref{firstpage}--\pageref{lastpage}} \pubyear{2007}
\maketitle
\label{firstpage}

\begin{abstract}

The distance to the relativistic jet source SS\,433 and the related
supernova remnant W50 is re-examined using new observations of H\,I in
absorption from the VLA, H\,I in emission from the GBT, and $^{12}$CO
emission from the FCRAO.  The new measurements show H\,I in absorption
against SS\,433 to a velocity of 75 km~s$^{-1}$ but not to the
velocity of the tangent point, which bounds the kinematic distance at
$5.5 \leq d_k < 6.5$ kpc. This is entirely consistent with a
$5.5\pm0.2$ kpc distance determined from light travel-time arguments
\citep{BlundellBowler}.  The H\,I emission map shows evidence 
of interaction of the lobes of
W50 with the interstellar medium near the adopted systemic velocity of
$V_{LSR} = 75$ km~s$^{-1}$.  The western lobe sits in a cavity in the
H\,I emission near the Galactic plane, while the eastern lobe
terminates at an expanding H\,I shell. The expanding shell has a
radius of 40 pc, contains $8\pm3 \times 10^3$ M$_{\sun}$ of H\,I and
has a measured kinetic energy of $3\pm1.5 \times 10^{49}$ ergs.  There
may also be a static H\,I ring or shell around the main part of W50
itself at an LSR velocity of 75 km~s$^{-1}$, with a radius of 70 pc
and a mass in H\,I of $3.5 - 10 \times 10^4$ M$_{\sun}$.  We do not
find convincing evidence for the interaction of the system with any
molecular cloud or with H\,I at other velocities. The H\,I emission
data suggest that SS\,433 lies in an interstellar environment
substantially denser than average for its distance from the Galactic
plane.

This Population I system, now about 200 pc below the Galactic plane,
most likely originated as a runaway O-star binary ejected from a young
cluster in the plane.  Given a modest ejection velocity of $\geq30$
km~s$^{-1}$, the binary could have reached its present location in
$\leq10$ Myr at which time the more massive member became a supernova.
New astrometric data on SS\,433 show that the system now has a
peculiar velocity of a few tens of km~s$^{-1}$ in the direction of the
Galactic plane.  From this peculiar velocity and the symmetry of the W50
remnant we derive a time since the SN of $\leq1 \times 10^5$ yr.

\end{abstract}
 

\section{Introduction}

From its discovery as an eclipsing X-ray binary star emitting
relativistic jets, the distance to SS\,433 and the W50 supernova
remnant (SNR) in which it is embedded has been somewhat uncertain.
The peculiar and exciting properties of this system were first
described by \citet{Margon79}, who estimated its kinematic distance as
3.5 kpc from observations of the velocity of highly saturated
foreground interstellar NaII lines and a straightforward model for
Galactic rotation.  As the properties of SS\,433 became better
understood --- it is a compact object in a binary system ejecting
relativistic jets along an axis 
which precesses through small angles --- the
determination of its distance became more important in fixing the
physical size of the observed phenomena.  From a  measurement of 
the proper motion  of individual radio structures within the SS\,433 jets,
and the assumption of a constant jet velocity, 
\citet{Hjellming81,Hjellming81a} derived a distance of
$5.5 \pm 1.1$ kpc, which was soon adopted as the canonical value for the system
\citep{Margon84}.

\begin{figure*}
%% figure 1
\centerline{
\includegraphics[width={0.75\textwidth}]{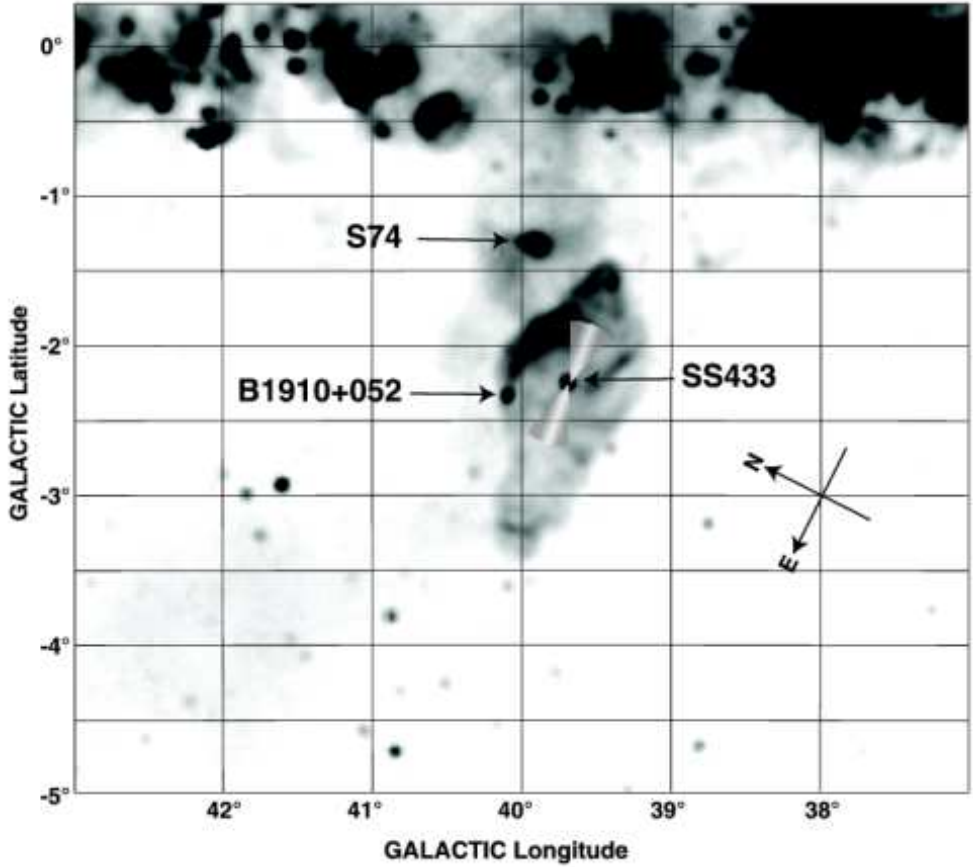}}
\caption{\label{schematic}The  11cm continuum image 
made at $4\farcm3$ angular resolution \citep{Reich} of the region
containing W50, which is centred on SS\,433.  The SS\,433
jets are shown schematically at the approximate correct orientation
and scale.  The H\,II region S74 and the extragalactic radio source
B1910+052, which is superimposed on the W50 shell, are also noted.}
\end{figure*}

Figure~\ref{schematic} shows SS\,433 and W50 in relation to the
Galactic plane and other objects in the same part of the sky.  SS\,433
is located at $\ell,b = 39\fdg7-2\fdg2$, in the centre of the
radio continuum source W50, which is thought to be either a 
supernova remnant or a stellar wind bubble \citep{Konigl}. 
 The SS\,433 system consists of a compact object
surrounded by an accretion disc and a `donor' star, of $M \gtrsim 10
M_{\sun}$ \citep{Hillwig}.  Material from
the donor star passes to an accretion disc surrounding the compact
object and generates two relativistic jets moving at 0.26c in opposite
directions which have apparently created the lobes (or ``ears'')
punched through the rim of the circular SNR W50 \citep{
Begelman,Margon84}.  The jet axis precesses with a cone opening angle
of $20\arcdeg$ around a position angle of $98\arcdeg$ \citep{Margon84,
Stirling}, though at their maximum extent they show evidence of having
been collimated to an effective opening angle $\sim 10\arcdeg$
\citep{Brinkmann2006}.  The jet closer to the Galactic plane is
usually referred to as the `Western' jet and the other as the
`Eastern'. The presence of a $\sim 10 M_{\sun}$ (hence young) star in
SS\,433 clearly marks this system as part of Population I, a
classification whose implications we will discuss in $\S7$.

The stellar component of SS\,433 cannot be assigned a distance from
traditional spectro-phometric or kinematic techniques: the nature of the 
donor star 
is still uncertain \citep{Fuchsetal}, as is the extinction, while the optical 
systemic radial velocity of $+82\pm3$ km~s$^{-1}$  
($V_{LSR}$) may have a significant random component acquired during 
the supernova (SN) \citep{Hillwig}. But as a radio source, SS\,433
can be used as a target for H\,I absorption measurements:  the
kinematics of the intervening gas seen in absorption can provide
limits on the location of SS\,433, limits which are not likely to be
affected by conditions close to the source itself
\citep{MurdinClarkMartin}.  Such measurements were first made by
\citet{vanG79} and showed that SS\,433, and by implication the entire
W50 remnant, absorbed H\,I only at a $V_{LSR} < 53$ km~s$^{-1}$, which
limits the kinematic distance to $d_k\geq3.7$ kpc.  The authors noted
that their measurements might be consistent with a kinematic distance
of as much as 4.7 kpc, but this rather uncertain upper limit was
rendered less probable by subsequent observations of H\,I absorption
toward the extragalactic radio continuum source B1910+052 only
$25\arcmin$ from SS\,433 (Fig.~\ref{schematic}), which has a rich
absorption spectrum at all velocities up to the maximum allowed from
Galactic rotation \citep{Dickey83}.  This implies that if SS\,433 were
actually 5 kpc from the Sun, H\,I should be seen in absorption against
it to at least 75 km~s$^{-1}$,  more than 20 km~s$^{-1}$
higher than was detected.  (Throughout this paper velocities are given
with respect to the LSR defined from ``Standard Solar Motion''
\citep{Delhaye} unless otherwise noted.)

Observations of $^{12}$CO from molecular gas added to the puzzle.
SS\,433/W50 lies more than 2 degrees from the Galactic plane, an
uncommonly large latitude for Population I objects in the inner Galaxy
except for those quite near to the Sun.  Observations of $^{12}$CO
showed that there is a large molecular cloud covering SS\,433 with a
velocity in the range 27--36 km~s$^{-1}$ and an implied distance of
2.2 kpc \citep{Huang, Yamamoto}, though the `far' kinematic distance
of $\sim 11$ kpc might also be possible.  The $^{12}$CO data
did not show strong kinematic or morphological evidence for an
interaction between W50 and the molecular cloud, but the coincidence of an
SNR and molecular cloud $\approx 2\arcdeg$ from the Galactic plane was
thought to make their association likely.  Furthermore, the HII region
S74 lies projected on the edge of the W50 remnant (Fig.~1), between
the remnant and the Galactic plane.  S74 is associated with $^{12}$CO
emission at 48 km~s$^{-1}$ and the spectro-photometric distance to its
exciting star is reported as $d_*=2.1 \pm 0.8$ kpc \citep{BrandBlitz,
Forbes}, though the quality of this distance determination is unclear.
At a distance of 2.1 kpc S74 is about 50 pc below the plane, well
within the range of displacements of HII regions in the inner Galaxy
\citep{LockmanPisanoHoward}.   SS\,433/W50 might thus plausibly be 
connected with S74, another Population I object, 
were they actually near each other 
along the line of sight.

The energetic event which produced the W50 remnant, and the action of
the SS\,433 jets, should leave some mark on the interstellar medium
(ISM), for most of the kinetic energy of the jets is not radiated
\citep{Brinkmann2006}. A study of the H\,I emission around W50 at
$21\arcmin$ angular resolution \citep{Dubner} reported evidence of an
interaction between W50 and the neutral ISM at velocities ranging from
35 to 50 km~s$^{-1}$.  \citet{Dubner} combined their data with
previous H\,I measurements toward W50 \citep{KooHeiles}, to propose a
model in which W50 has a systemic velocity of $V_{LSR} = 42$
km~s$^{-1}$ and is surrounded by a shell of H\,I expanding at $\pm 76$
km~s$^{-1}$.  The kinematic distance proposed for SS\,433/W50 by Dubner
et al.~is $d_k = 3$ kpc.

Recently a new distance estimate for SS\,433 has been derived 
by \citet{BlundellBowler}, who fitted a 
high-resolution radio observation of the SS\,433 jets 
to a projection of the standard kinematic model over two complete precession 
cycles of the jets, and
analyzed asymmetries in the 
radio structure over arc-second scales caused by light-travel time 
effects.  They 
find a distance of $5.5\pm0.2$ kpc, which differs by as much as 
 a factor $\sim2$  from  distance estimates based on 
the association between SS\,433/W50 and interstellar matter.

The discrepancy 
has prompted us to obtain new H\,I 
absorption, H\,I emission and $^{12}$CO measurements of the
SS\,433/W50 system to reassess the  contradictory evidence.  In
this paper we  discuss the Galactic kinematics in the direction of
SS\,433, present the new H\,I and CO data, then show that they are entirely
consistent with a distance of 5.5\,kpc for the system.  
We also find evidence in the new H\,I emission data 
 for interaction between SS\,433/W50 and the ambient 
ISM, for a static ring or H\,I shell around W50,  and for 
an expanding H\,I shell which may be 
 `fossil' evidence that the Eastern jet of  SS\,433 once extended 
past the current boundries of the radio continuum source.

\section{Observations}

\subsection{VLA H\,I Absorption Measurements}

The Very Large Array (VLA) of the NRAO was used to observe SS\,433
during two periods of about 40 minutes each on 5 June 1998 when it
was being moved from A to B configuration. The nearby 1.2-Jy object
J1950+081 was used as a phase calibration source. The bandpass and
total flux density scale were calibrated by observations of 3C48.  The
first observations were made with a velocity coverage of
160\,km\,s$^{-1}$ at a resolution of 1.29 km~s$^{-1}$ over 127
channels. The 
synthesized beam is $2\farcs56 \times 1\farcs54$ at a position angle
of $90\arcdeg$. The rms noise per channel is\,3.5 mJy/beam while the
continuum peak is 612\,mJy/beam. During the second set of independent
observations the bandwidth was twice as large, giving a velocity
resolution of 2.58\,km~s$^{-1}$, an rms noise of 3.0\,mJy/beam, and
angular resolution of $3\farcs43 \times 1\farcs59$ at a position angle
of $76\arcdeg$ for this epoch. The continuum peak at this resolution
is 750\,mJy/beam. The rms noise in H\,I opacity per channel is about
0.004 for both sets of observation.

\subsection{GBT H\,I Emission Measurements}

An image in the 21-cm emission line of H\,I was constructed for the 
 region around SS\,433 using archival data from the Robert C. Byrd 
Green Bank Telescope (GBT) as well as observations made 
 specifically for this project with the GBT.  In all cases the spectra 
cover about 500\,km~s$^{-1}$ 
in total, centred at $+50$ km~s$^{-1}$ with a channel spacing of 
1.03\,km~s$^{-1}$ and an effective velocity resolution of 1.25\,km~s$^{-1}$.  
The region around W50 was mapped by making a series of scans 
 in Galactic longitude at a fixed latitude.  Spectra were measured 
every $3\arcmin$  in both coordinates. Each position was observed for $4^s$.  
The data were  gridded to a final resolution of 
$3\farcm5$, slightly finer than Nyquist sampling for the $9\farcm1$
angular resolution of the GBT.  The `on-the-fly' observing and subsequent 
gridding gave the final maps an effective angular resolution of 
$10\farcm2 \times 9\farcm6$ in Galactic longitude and latitude, respectively.
 A 2nd order polynominal was 
fitted to emission-free regions of the spectra,  and 
the rms noise in the final spectra is 0.10 K.
  
The H\,I emission associated with SS\,433 is relatively bright, so
stray radiation from GBT sidelobes \citep{LockmanCondon} should not be
an issue here.  There was, however, a small error in the velocity
tracking while observing some portions of the mapped region, which
resulted in an $\sim0.1$ km~s$^{-1}$ error in the velocity scale of
many spectra.  This is $\lesssim10\%$ of a velocity channel width, but
it can vary systematically from row to row, and can be seen as a faint
striping when there are strong gradients in $T_b(V)$.  For
quantitative purposes this problem is negligible, but it does occur in
some of the images.

\begin{figure}
%% figure 2
\centerline{
\includegraphics[width={0.5\textwidth}]{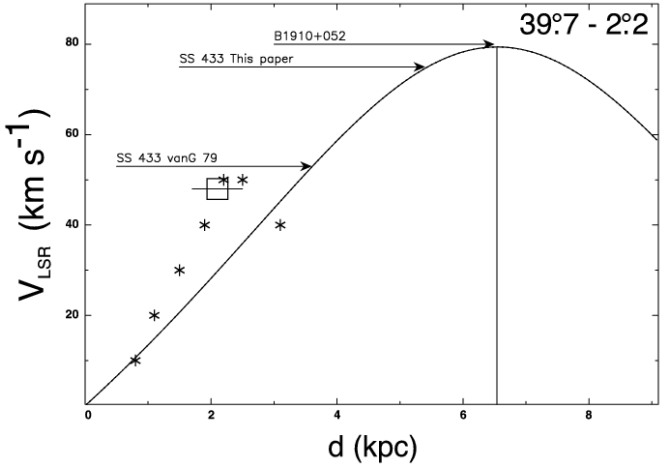}}
\caption{\label{vofr} Relationship between $V_{LSR}$ and distance from the Sun 
in the direction of SS\,433 for a flat rotation curve with $V_0 = 220$
km~s$^{-1}$.  A vertical line is drawn at the tangent point distance.
The rectangle at 2.2 kpc, 48 km~s$^{-1}$, marks the velocity and
distance of the HII region S74; the horizontal bar covers a 20\%
uncertainty in distance.  The asterisk symbols show the Galactic
velocity field derived by \citet{BrandBlitz} from distances to the
exciting stars of HII regions in the general direction of SS\,433.  
Horizontal arrows are drawn at the maximum extent of the H\,I absorption
measured toward SS\,433 by \citet{vanG82} and our current mesurements.
The intersection of the arrow with the velocity curve gives the
kinematic distance.  For comparison, the maximum velocity of H\,I
absorption toward the extragalactic radio source B1910+052, which lies
about $25\arcmin$ from SS\,433, is also shown \citep{Dickey83}.  The
new data suggest that SS\,433 has a kinematic distance $d_k \geq 5.5 $
kpc, and that it is not as distant as the tangent point.  }
\end{figure}

\subsection{FCRAO $^{12}\rm{CO}$ Emission Map}

The $^{12}$CO data were obtained for us by C.~M.~Brunt and M.~H.~Heyer
using the 14 m telescope of the Five College Radio Astronomy
Observatory (FCRAO) with the SEQUOIA 32 pixel focal plane array.  An
area about $2\arcdeg \times 2\arcdeg$ centred on SS\,433 was fully
sampled at an angular resolution of $45\arcsec$, and the data were
gridded into a cube with a pixel size of $20\arcsec$. Spectra cover
65\,km~s$^{-1}$ from 7.5 to 72.5\,km~s$^{-1}$ at a velocity resolution
of 0.063\,km~s$^{-1}$, though for this paper we smoothed the spectra
to a velocity resolution of 0.25\,km~s$^{-1}$.  The rms noise in the
spectra is about 0.4 K.  Further information on the FCRAO 14-m system
is given in \citet{Heyer}.

\section{H\,I Absorption and the Kinematic Distance of SS\,433}

Figure~\ref{vofr} shows the expected run of $V_{LSR}$ with distance
from the Sun, $d$, in the direction of SS\,433 for a flat rotation
curve with $R_0=8.5$ kpc and $V_0 = 220$ km~s$^{-1}$.  Other models
for Galactic rotation in the inner Galaxy (e.g., \citet{Burtonrot,
Clemens}) differ from a flat curve by only a few km~s$^{-1}$ at the
distances shown in the Figure.  For an azimuthally symmetric rotation
curve, $V_{LSR}$ is symmetric about the tangent point distance, $d_t =
R_0 \ \cos(\ell)$, then becomes negative at $d>13 $ kpc, where 
 $R >R_0$. The value of $V_{LSR}$ at the tangent point, $V_t$, for the flat
rotation curve model is within a few km~s$^{-1}$ of measured values of
$V_t$ for $^{12}$CO in the Galactic plane at similar longitudes
\citep{Clemens} and also agrees with the maximum velocity of H\,I 
absorption toward the extragalactic radio source B1910+052, which lies
projected on the rim of W50 \citep{Dickey83}.  

The line of sight to SS\,433 at $\ell \approx 40\arcdeg$ lies beyond
the region of the Galaxy influenced by strong streaming motions around
the Galactic bar (e.g., \citet{WeinerSellwood}), but large-scale
density wave motions may still be important (e.g.,
\citet{EnglmaierGerhard06}).
\citet{BrandBlitz} have tried to determine the true velocity 
field of the Galaxy using distances to the exciting stars of HII
regions and velocities from $^{12}$CO measurements of the associated
molecular clouds.  Their results in the direction of SS\,433 are shown
by starred symbols in Fig.~\ref{vofr}.  The HII region S74 (which is 
projected on the edge of W50) is shown as an open square, with the
$20\%$ distance uncertainty estimated by
\citet{Forbes}.  The S74 distance is based on measurement of a single
star and may have an error larger than $20\%$.  S74 itself has
significant weight in the determination of the Brand \& Blitz velocity
field in this direction, but there are a few other HII regions whose
distance and kinematics also imply that the Galactic velocity field
rises somewhat steeply for a few kpc from the Sun before returning to
the `flat' value at $d = 3$ kpc
\citep{BrandBlitz}.

The arrow marked `SS\,433 vanG~79' shows the maximum velocity of H\,I
absorption toward SS\,433 found by \citet{vanG79}.  While this gives
$d_k \geq 3.7$ kpc for a flat rotation curve, the \citet{BrandBlitz}
empirical velocity field suggests that the limit might be even less
stringent: $d_k \gtrsim 2$ kpc.  The arrow labelled `B1910+052' marks
the velocity limit of H\,I absorption toward the radio continuum
source B1910+052 only $25\arcmin$ away from SS\,433
\citep{Dickey83}. This source has an H\,I absorption component at -20
km~s$^{-1}$ showing that it is $>10$ kpc from the Galactic Centre and
thus almost certainly extragalactic. At positive velocities it has
H\,I in absorption continuously between 0 and 80 km~s$^{-1}$ and
supplies us with two important facts: cool H\,I is most likely present
at all permitted velocities along the line of sight to SS\,433, and the
value of $V_t$ given by the flat rotation curve is correct to within a
few km~s$^{-1}$.

\begin{figure}
%% figure 3
\centerline{
\includegraphics[width={0.5\textwidth}]{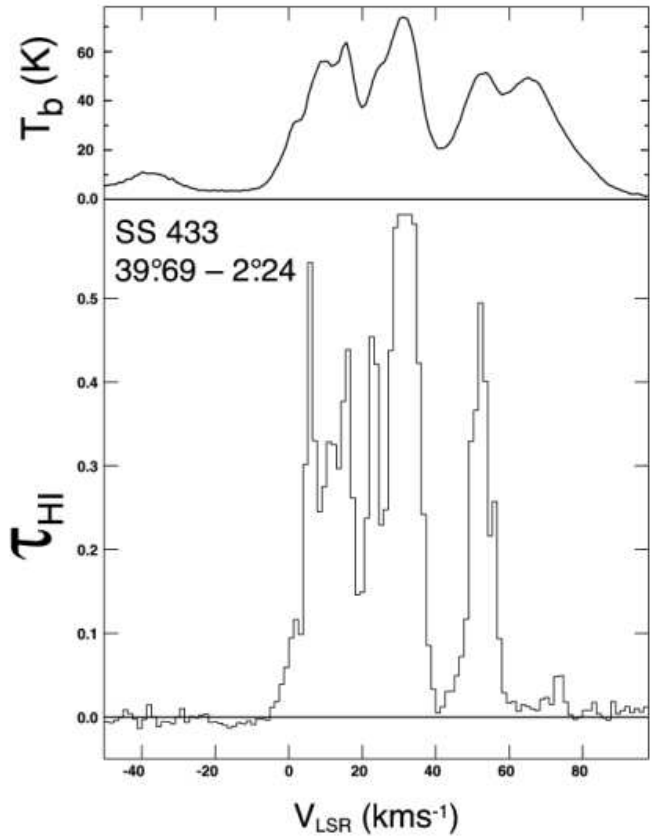}}
\caption{\label{SS433HI} The upper 
panel shows the GBT 21cm H\,I emission spectrum towards SS\,433 and 
the lower panel shows the high-velocity resolution VLA absorption
spectrum plotted as $\tau(V_{LSR})$.  Significant H\,I absorption at 75 
km~s$^{-1}$ is seen for the first time and establishes the kinematic
distance of SS\,433 as $\geq5.5$ kpc.}
\end{figure}

Figure~\ref{SS433HI} shows the spectrum of H\,I in absorption against SS\,433
(lower panel) obtained from the new high-velocity resolution VLA
observations and the corresponding emission spectrum from the GBT
observations (upper panel).  The independent VLA absorption spectrum
made with lower velocity resolution is consistent with the data shown
here.  H\,I absorption components are detected to $V_{LSR} = 
75$ km~s$^{-1}$, a kinematic distance of $d_k \geq 5.5$
kpc, as indicated by the arrow in Fig.~\ref{vofr}.  The relatively weak
$\tau_{HI} = 0.05$ absorption at $V_{LSR} = 75$ km~s$^{-1}$ was not
detected in previous data, though hints of it at the $3\sigma$ level
can be seen in the data of \citet{Dickey83}.  In the new VLA spectra
it is significant at a level $>10\sigma$ above the noise.

\begin{figure}
%% figure 4
\centerline{
\includegraphics[width={0.5\textwidth}]{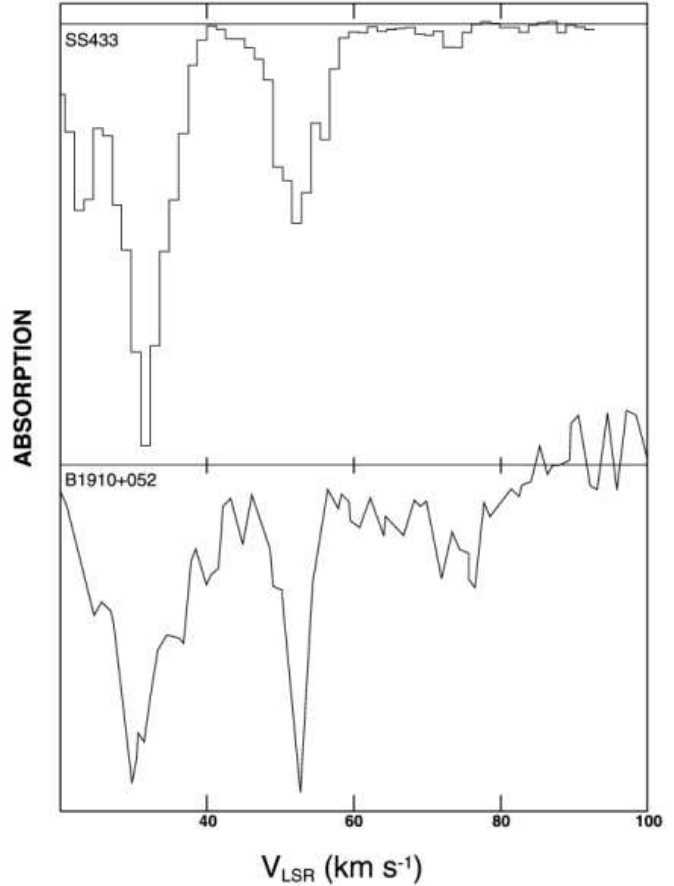}}
\caption{\label{HIabs} 
The high-velocity portion of the new VLA H\,I absorption spectrum
toward SS\,433 (upper panel) compared with that from \citet{Dickey83}
toward B1910+052, an extragalactic continuum source only $25\arcmin$
away from SS\,433 at a similar Galactic latitude (lower panel). The
absence of H\,I absorption at $V_{LSR} \sim 80$ km~s$^{-1}$ in the
spectrum toward SS\,433, and its presence in the spectrum toward the
extragalactic source, implies that SS\,433 lies in front of the tangent
point and thus at $5.5 \leq d_k < 6.5$ kpc.}
\end{figure}

Figure~\ref{HIabs} shows the high-velocity portion of the VLA H\,I
absorption spectrum toward SS\,433 (this paper) and toward B1910+05, the
extragalactic source only $25\arcmin$ from SS\,433
\citep{Dickey83}.  The absence of H\,I absorption toward SS\,433 at
velocities $\gtrsim 80$ km~s$^{-1}$,  which is the terminal velocity 
where absorption is seen toward
B1910+05, implies that SS\,433 is nearer than the tangent point
distance $d_t = R_0 \ \cos(\ell) = 6.5$ kpc.  Thus the H\,I absorption
data taken in total give $5.5 \leq d_k < 6.5$ kpc.  

These values are consistent with the distance derived from a
light-travel time analysis and we therefore adopt the
\citet{BlundellBowler} distance of $d = 5.5\pm0.2$ kpc, and take
$V_{LSR} = +75\pm6$ km~s$^{-1}$ as the systemic velocity of the the
equivalent standard of rest of the SS\,433/W50 system, where the
uncertainty is the typical random motion of cool H\,I clouds
\citep{DickeyLockman}.  This velocity is close to the mean velocity of
the SS\,433 binary system \citep{Hillwig}, which, however, may have a
significant peculiar velocity acquired during the SN event.  This is
discussed more fully in $\S7$.

\section{SS\,433 and the Molecular Cloud at 30 km~s$^{-1}$}

In this section we examine the evidence that SS\,433 is interacting with 
a molecular cloud at a velocity of about 30 km~s$^{-1}$, as 
has been suggested by 
some previous work \citep{Huang, Yamamoto}.  

\begin{figure}
%% figure 5
\centerline{
\includegraphics[width={0.5\textwidth}]{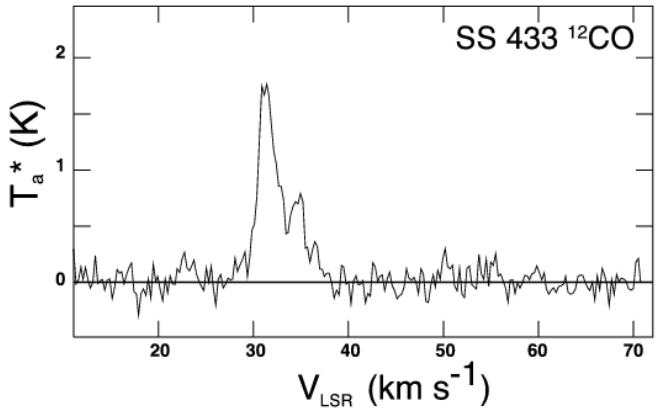}}
\caption{\label{COspect} A $^{12}$CO spectrum 
from the FCRAO 14 meter telescope toward SS\,433.  For this Figure the 
data have been smoothed to $1\arcmin$ angular resolution and 0.25 km
s$^{-1}$ velocity resolution. The intensity units are $T_a^*$.}
\end{figure}

\begin{figure}
%% figure 6
\centerline{
\includegraphics[width={0.5\textwidth}]{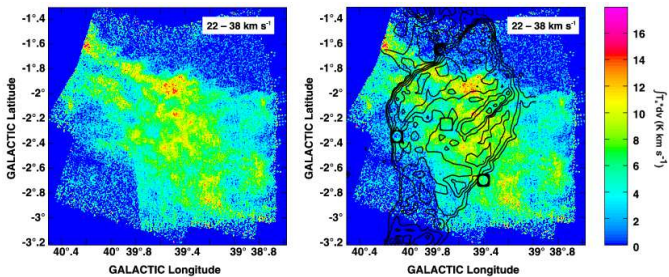}}
\caption{\label{COmaps} The $^{12}$CO data from the FCRAO 14m telescope 
integrated between 22 and 38 km~s$^{-1}$, a range which includes
almost all of the emission in the field.  The intensity units are
K-km~s$^{-1}$ of $T_a^*$.  The right panel has contours of 1.4 GHz
continuum from \citet{Dubner} atop the $^{12}$CO to show the location
and extent of SS\,433 and W50.  The $^{12}$CO emission does not bear
any morphological resemblence to the continuum features, and there is
no evidence we can find in the data which suggests that there is
interaction between the remnant or jets and the molecular cloud.}
\end{figure}

Figure~\ref{COspect} shows a $^{12}$CO spectrum towards SS\,433 derived
from the new observations with the FCRAO, while Figure~\ref{COmaps}
shows the $^{12}$CO emission integrated over the velocity range 22--38
km~s$^{-1}$.  The image includes $^{12}$CO at all velocities which
were considered to be connected with W50 in the earlier studies
\citep{Huang, Yamamoto}, which is virtually all of the $^{12}$CO in
this part of the sky.  The right panel of Fig.~\ref{COmaps} shows the
integrated $^{12}$CO intensities overlaid with contours of the 1.4 GHz
continuum of W50 from \citet{Dubner}.  There is no obvious
relationship between the $^{12}$CO emission and the radio continuum,
and, except that they both lie somewhat below the Galactic plane,
nothing in these data suggests any connection.  Scrutiny of  line
velocity and line width maps leads to a similar conclusion. 

The FCRAO $^{12}$CO data also show that the S74 HII region 
(Fig.~\ref{schematic}) is 
 not connected with either the SS\,433/W50 system, or with the 
molecular cloud at 22 -- 38 km~s$^{-1}$.  
$^{12}$CO emission associated with S74 is seen
only between 40 and 47 km~s$^{-1}$, and only in the immediate 
vicinity of the nebula. 
   There is no $^{12}$CO at the velocity of S74 within the 
boundries of W50,  so the HII region 
S74 seems also to be  a chance
overlap on the sky with the edge of the remnant.  

The highest velocity $^{12}$CO emission in this field comes at 56
km~s$^{-1}$ from a small cloud at $\ell,b = 30\fdg6-1\fdg6$, just
touching the top of the western lobe.  Both the FCRAO and the
lower-resolution $^{12}$CO data \citep{Huang, Dame} have no
significant $^{12}$CO emission at velocities $>56$ km~s$^{-1}$ except
for general emission from the molecular cloud layer near the Galactic
plane at $b \geq -1\arcdeg$. Thus we find no evidence for a molecular
cloud near the longitude, latitude and $V_{LSR}$ of SS\,433. This has
implications for the origin of the system, and will be addressed in
$\S7$.

In summary, despite their overlap on the sky, there is no evidence in
$^{12}$CO for an association between the molecular gas near 30
km~s$^{-1}$ and the SS\,433/W50 system.

\section{The Interaction Between  SS\,433/W50 and Interstellar H\,I}

\subsection{The Western Jet: A Cavity in H\,I}

\begin{figure}
\centerline{
\includegraphics[width={0.5\textwidth}]{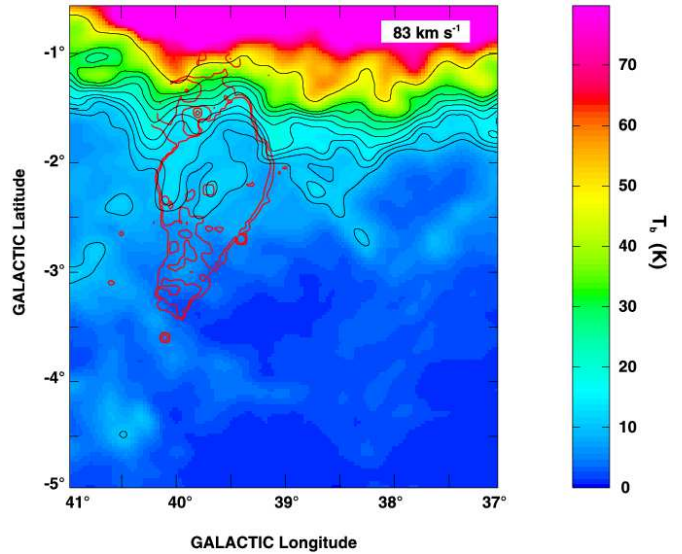}}
\caption{\label{ringv83}The GBT 21cm H\,I emission image 
at 83 km~s$^{-1}$ showing evidence for the interaction between the
upper part of the W50 remnant including the western jet of SS\,433, and
H\,I near the Galactic plane.  H\,I contours are plotted every 2 K
from 10~K to 16~K, every 4 K from 20~K to 32~K, and at 48~K. The
location of W50 in the 1.4 GHz radio continuum \citep{Dubner} is shown
in red. The faint H\,I feature near $40\fdg5$ --4$\fdg5$ lies along 
the axis of the eastern jet and is shown in more detail in Fig.~8.}
\end{figure}

The Western lobe of SS\,433 terminates at $b \approx -1\fdg5$ where the
W50 radio continuum contours sit in a slight cavity in the H\,I, seen
most clearly at $V_{LSR} > 75$ km~s$^{-1}$, e.g., at the 83
km~s$^{-1}$ emission shown in Figure~\ref{ringv83}.  It is likely that
the interaction is more visible at the higher velocities simply
because there is less confusion from unrelated emission.

%% figure 7

\subsection{The Eastern Jet: An Expanding H\,I Shell}

\begin{figure*}
\centerline{
\includegraphics[width={0.95\textwidth}]{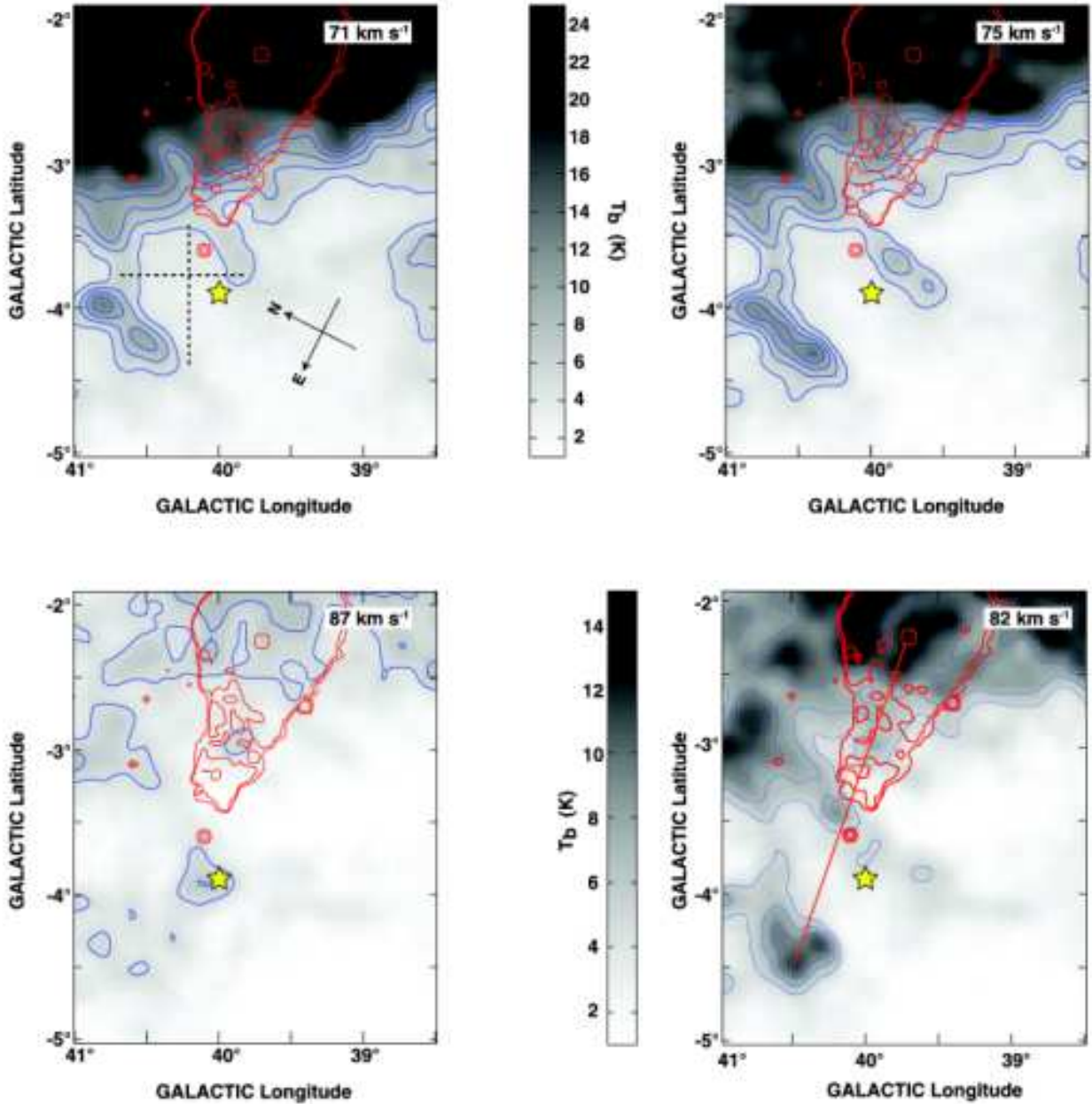}}
\caption{\label{lb_bubble} Several channel maps 
from GBT observations showing the expanding H\,I shell at 
$40\fdg2$--3$\fdg9$ in relation to the eastern end of W50.  
It has a velocity and
orientation coincident with that of the SS\,433/W50 system. The solid
red line in the lower right panel lies along the extension of the axis
of symmetry of the SS\,433 jets.  The dashed horizontal and vertical
lines in the upper left panel show the locus of the velocity-longitude
and velocity-latitude cuts displayed in Figure~\ref{vlvb_bubble}.  The
H\,I spectrum shown in Fig.~\ref{bubblespect} was taken at the
location marked by the star. H\,I contours are at 5, 7, 9 and 11 K.
The greyscale wedge in the upper centre applies to all but the lower
right panel, which has its own calibration wedge, but identical
contour levels.  }
\end{figure*}

The eastern lobe of the continuum source terminates abruptly at the
edge of a well-defined expanding shell of H\,I, shown at several
velocities in Figure~\ref{lb_bubble}. The shell is at its maximum
angular extent at the velocities of the top two panels of the Figure,
and at its greatest velocity in the lower left panel, where it has
contracted to a single receding `cap'.  The expanding shell is 
 open to the South, perpendicular to the axis of the jet, and may
also have a gap where it overlaps the continuum lobe itself. The solid
red line in the Figure lies along the extension of the axis of the
SS\,433 jets.  At 82 km~s$^{-1}$ the axis of the jet intersects a
chevron of H\,I which points in the direction of the jet motion (lower
right panel).  This feature is also visible in Fig.~\ref{ringv83}.

%% figure 8

%% figure 9
\begin{figure*}
\centerline{
\includegraphics[width={0.85\textwidth}]{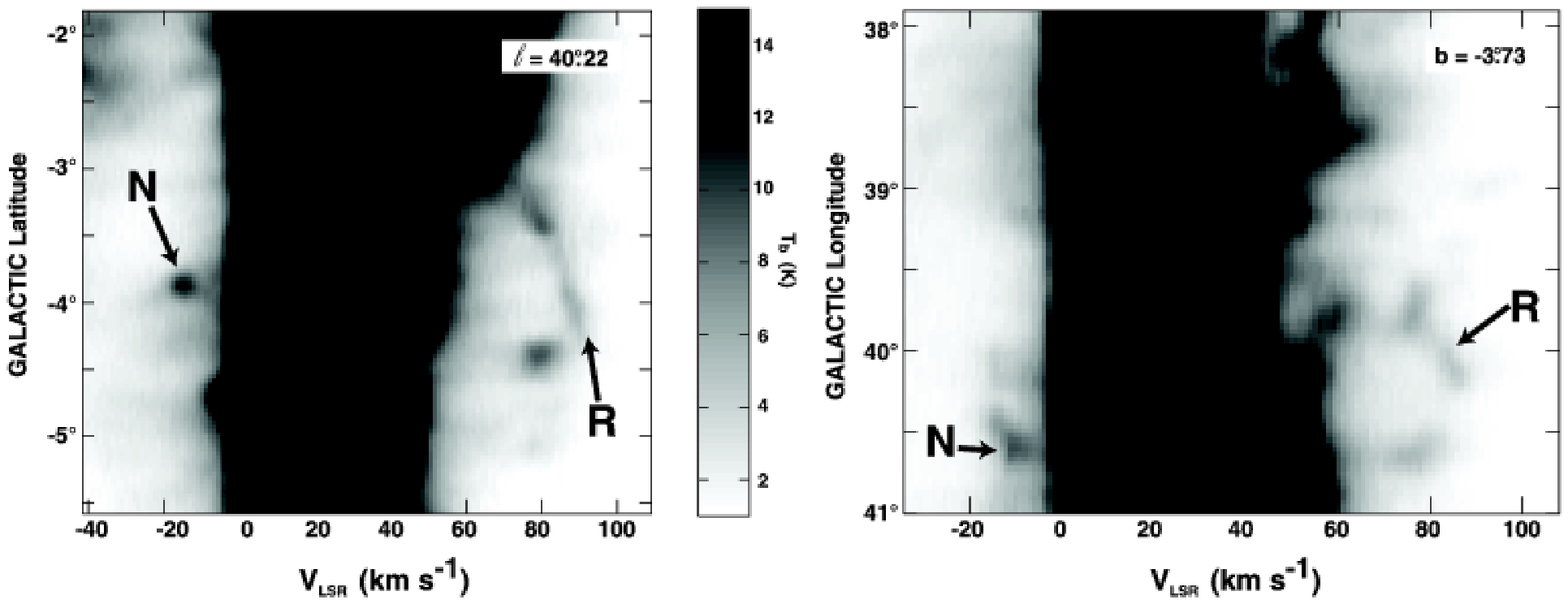}}
\caption{\label{vlvb_bubble} GBT H\,I data in 
velocity-longitude and velocity-latitude 
 cuts at $\ell = 40\fdg22$ and $b = -3\fdg73$ 
 showing the expanding shell  at 
$\ell, b, V_{LSR} = 40\fdg5, -3\fdg9, +71 $ km~s$^{-1}$.  
This structure extends
over $39\fdg6 < \ell < 40\fdg8$ and 
$-4\fdg0 < b    < -3\fdg4$, has 
an expansion velocity of $\pm16$ km~s$^{-1}$, and
touches the tip of the lower W50 lobe.  The letter `R' identifies the
receding part of the shell, and `N' a negative-velocity H\,I feature
which might be associated with it.  }
\end{figure*}

\begin{figure}
\centerline{
\includegraphics[width={0.5\textwidth}]{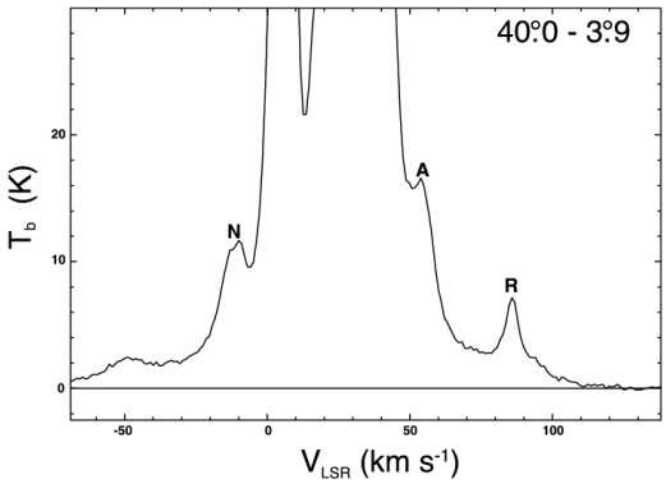}}
\caption{\label{bubblespect}The GBT H\,I spectrum 
through the expanding shell at 
$40\fdg0$--3$\fdg9$, the position marked by a star in the 
panels of Fig.~\ref{lb_bubble}.  Spectral components associated with the 
receding and approaching side of the bubble are indicated with 
`R' and `A'.  There is  a possible negative velocity 
component `N' which will be discussed in a later section. 
}
\end{figure}

The expansion of the shell is demonstrated in
Figure~\ref{vlvb_bubble}, through velocity-longitude and
velocity-latitude cuts along the direction of the dashed lines marked in
Fig.~\ref{lb_bubble}.  The centre of kinematic symmetry of the shell
is at $40\fdg2$--$3\fdg9$ at $V_{LSR}\approx 71$ km~s$^{-1}$, and its
expansion velocity is $V_{ex} = \pm 16$ km~s$^{-1}$.
Figure~\ref{bubblespect} shows a GBT H\,I spectrum at
$40\fdg0$--$3\fdg84$, the position marked by the yellow star in
Fig.~\ref{lb_bubble}.  This particular spectrum was chosen for display
because it shows the major components of the system most distinctly.
The receding part of the shell (marked `R') is well-separated in
velocity from other H\,I emission and reaches $V_{LSR} \gtrsim 85$
km~s$^{-1}$, well beyond the terminal velocity in its direction
(Fig.~\ref{vofr}).  The approaching side of the shell, labelled A, is a
distinct component in the spectra, though it cannot be measured
accurately at most locations. In this direction the receding component
has $N_{\rm HI}=4.9 \times 10^{19}$ cm$^{-2}$ and a line width $\Delta
V = 6$ km~s$^{-1}$ (FWHM), values typical of directions through the
centre of the shell, which have a range in $N_{HI}$ of 3--9 $\times
10^{19}$ cm$^{-2}$ and in $\Delta V$ of 4--10 km~s$^{-1}$.  In many
places the shell appears to be barely resolved by the GBT beam: it
must have an intrinsic thickness (FWHM) $\lesssim 12\arcmin$, or
$\lesssim 20$ pc at the distance of SS\,433.

%% figure 10

\subsubsection{Mass, density, and energetics}

At an adopted distance of 5.5 kpc, a simulated  shell with a 
Gaussian radial density profile, a radius to the peak of 40 pc 
and a thickness (FWHM) of 20 pc fit the data reasonably well.  
Assuming that the gap in the shell covers $10\%$ of its surface 
area, the total H\,I mass is $8 \times 10^3  M_{\sun}$.  
For the expansion velocity of $\pm16$ km~s$^{-1}$, and a mean particle mass 
of $1.4m_H$, the total measured kinetic energy in the shell is 
$E_{\rm k} = 3 \times 10^{49}$ ergs.     The  H\,I mass divided by the volume 
of the sphere to its outermost radius  gives the 
initial density $n_0$ = 0.4--1.2 cm$^{-3}$, a range which covers all 
plausible values of its size.  This is about an order of magnitude 
larger than the interstellar densities 
expected this far from the plane \citep{DickeyLockman}. 
The properties of the expanding shell are 
 summarised in Table~1, where the errors attached to each quantity include 
the range of measurement uncertainty and the effect of a 
$\pm1$ kpc change in the adopted distance.

\subsection{A static ring or shell around W50}

\begin{figure}
\centerline{
\includegraphics[width={0.5\textwidth}]{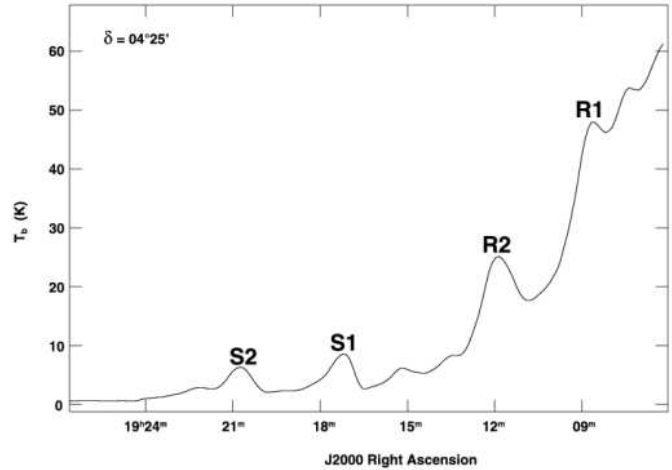}}
\caption{\label{Deccut} The H\,I brightness temperature 
at $V_{LSR} = 75$ km~s$^{-1}$ from GBT data 
along a cut at $\delta = 04\arcdeg25\arcmin$, 
approximately parallel to the jets of SS\,433 but 
$35\arcmin$ south of SS\,433 itself.  This slice shows four H\,I ridges
possibly associated with SS~443/W50. S1 and S2 are cuts through the
expanding shell (Fig.~8).  R1 and R2 are part of what appears to be an H\,I
ring or static shell which surrounds the western part of W50. The
strong H\.\,I brightness gradient to lower Right Ascension results
from the change in Galactic latitude from -$5\arcdeg$ to -$1\arcdeg$
over this range. }
\end{figure}

The H\,I emission at the adopted systemic velocity of 75 km~s$^{-1}$,
as measured with the GBT along a path at constant declination,
approximately parallel to the long axis of the SS\,433 system, is shown
in Figure~\ref{Deccut}.  This track cuts through four distinct ridges
of H\,I which are superimposed upon a strong gradient.  The two ridges
at higher Right Ascension belong to the expanding shell discussed in
the previous section. The two ridges at lower Right Ascension appear
to be the edges of a static H\,I ring or shell.

%% figure 11

\begin{figure*}
\centerline{
\includegraphics[width={0.50\textwidth}]{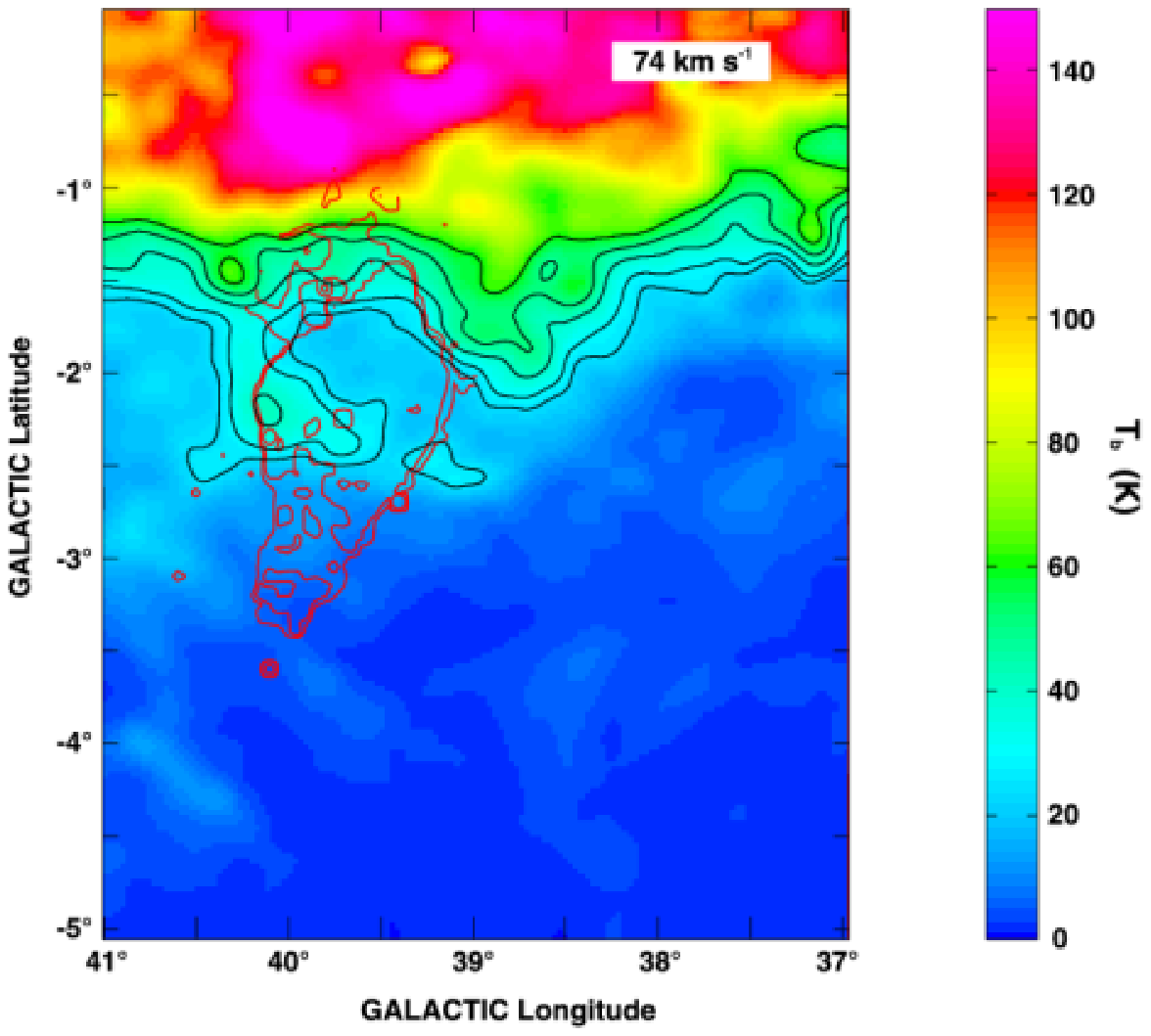}
\hspace{0.1in}
\includegraphics[width={0.50\textwidth}]{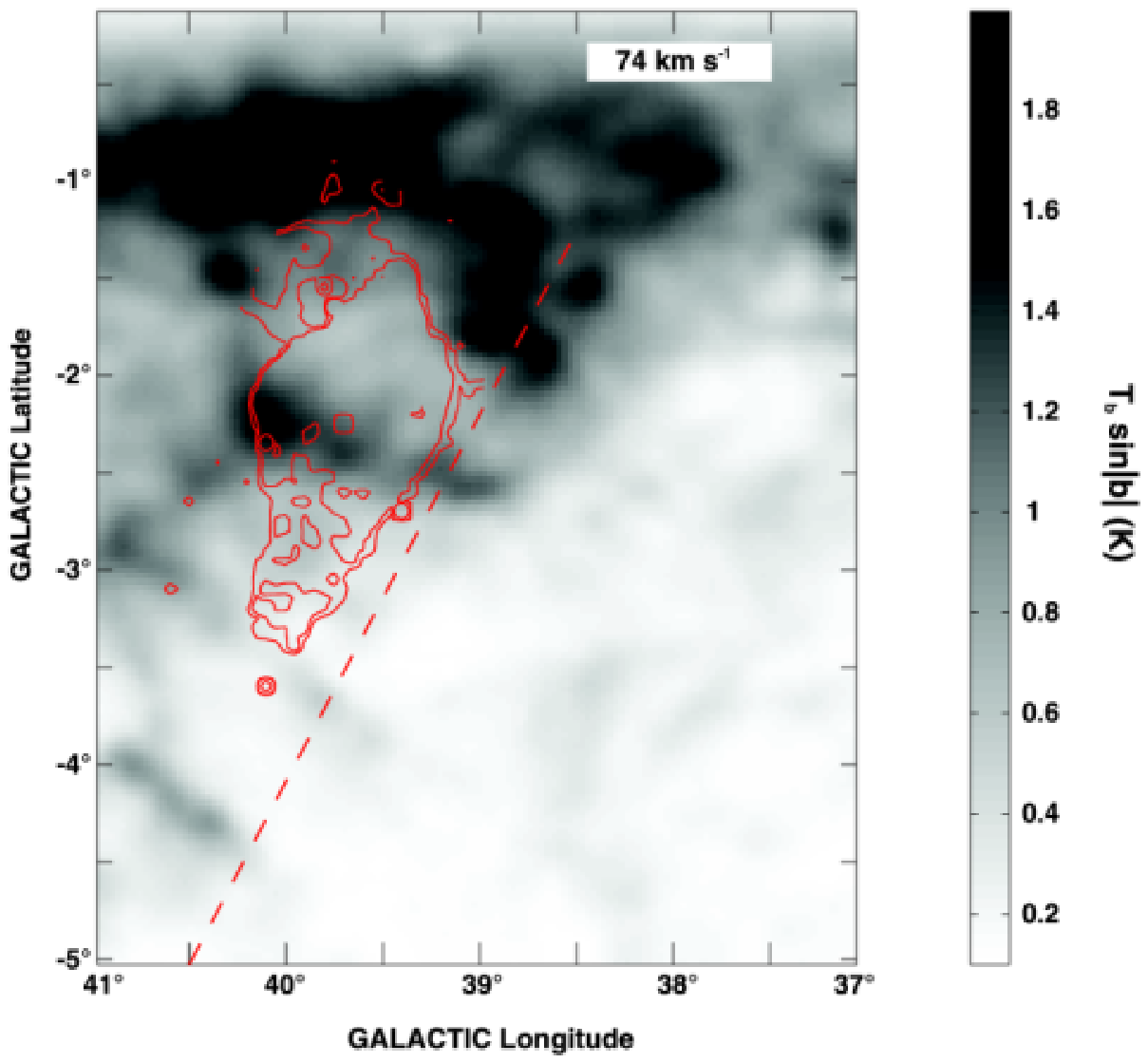}}
\caption{\label{ringv74} The GBT 21cm H\,I emission in a 
2 km~s$^{-1}$ velocity range around 74 km~s$^{-1}$ showing the static
H\,I ring surrounding the upper part of the W50 system.  The image on
the left shows H\,I brightness temperature while the image on the
right has temperatures scaled by $\sin|b|$ to increase the dynamic
range and accentuate deviations from a plane-parallel H\,I layer. The
outline of the 1.4 GHz radio continuum from W50 \citep{Dubner} is in
red; SS\,433 is at the central red contour.  The location of the
1-dimensional cut of Figure 11 is shown by the dashed red line.  }
\end{figure*}

The GBT image of H\,I emission at 74 km~s$^{-1}$ is shown in
Figure~\ref{ringv74}.  In the right
panel the temperatures have been scaled by $\sin|b|$ 
to flatten somewhat the very strong H\,I brightness gradient
near the Galactic plane and accentuate H\,I features which have a
significant deviation from a smooth layer.  The two western ridges in
Fig.~\ref{Deccut} come from cuts through an irregular ring of
emission centred at $\ell,b = 39\fdg6 -1\fdg8$ on the upper part of
the W50 remnant.  It is possible that the H\,I cavity illustrated in
Fig.~\ref{ringv83} is part of this ring.

%% figure 12

\begin{figure}
\centerline{
\includegraphics[width={0.5\textwidth}]{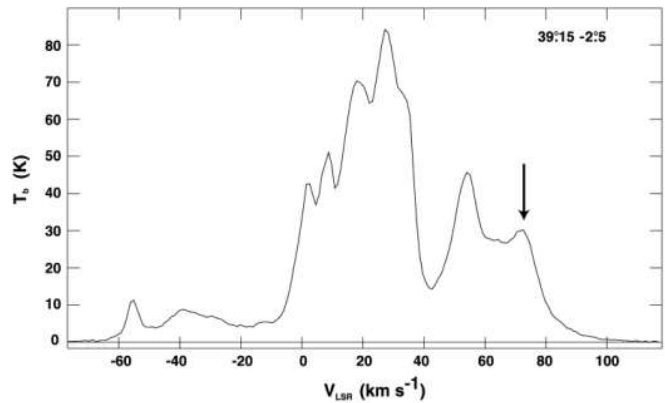}}
%\epsscale{0.95}
%\plotone{/users/jlockman/blundell/paper/figures/final_ringspect.eps}
\caption{\label{ringspect}The GBT 21cm H\,I emission spectrum toward a 
$39\fdg14$--2$\fdg50$, a position on the lower part of the static H\,I
ring around W50.  The ring component is the highest velocity peak in
the spectrum, marked with the arrow.}
\end{figure}

The H\,I ring appears to be slightly elliptical with an average
diameter $ \approx135$ pc (for a distance of 5.5 kpc) and a thickness
in places of about 40 pc.  Figure~\ref{ringspect} shows the H\,I
spectrum at $39\fdg15$ $-2\fdg50$ through the lower part of the ring.
The ring is the highest velocity feature in the spectrum, severely
blended with lower velocity gas, but distinguishable over the range
$65 \lesssim V_{LSR} \lesssim 84$ km~s$^{-1}$.  Its linewidth $\Delta
V \sim 16$ km~s$^{-1}$ (FWHM) and the brighter portions have $N_{HI} =
5.5 \times 10^{20}$ cm$^{-2}$.

%% figure 13

We can find no evidence for expansion of this object.  While an
approaching segment might easily be lost in confusion from unrelated
emission, any receding segment should be detectable 
easily in the wings of the spectra (see also $\S6.2$). 
 The absence of evidence for expansion
suggests either that the structure is a ring or torus, not a complete
shell, or that it is a stalled shell with an expansion velocity less
than a few km~s$^{-1}$.

\subsubsection{Mass and Density}

The uncertain topology causes a considerable 
uncertainty in the derived mass because of the need to disentangle the 
object from unrelated emission. 
The visible ring  contains $\sim 3.5 \times 10^4\  M_{\sun}$ of H\,I, 
assuming the centre of the ring is at the background level. 
It can also be modelled as a complete shell, however, 
whose ring-like appearance 
results mainly from limb-brightening.  In this case the data can be 
matched by a shell with a Gaussian radial density profile with a peak 
H\,I volume density $n_{HI}=1.2$ cm$^{-3}$ at a radius of 80 pc,  
 a FWHM of 40 pc, and a total  mass in H\,I 
of $10^5  M_{\sun}$.  The average interstellar density is given by the 
mass divided by the volume, which, because the volume of this object depends 
on its assumed shape, is not well constrained.  Over the 
 range of plausible assumptions 
we find $n_0 \approx 1-2$ cm$^{-3}$, 
at least an  order of magnitude greater than 
the average H\,I density expected this far from the Galactic 
plane \citep{DickeyLockman}, a relative excess similar to that found 
around the expanding shell.

\section{H\,I emission Toward SS\,433/W50 at other velocities}

We searched for a possible association between W50 and H\,I emission at 
all other velocities and present the results here.  Although there 
are some coincidences, we do not believe that they are significant.

%% figure 14
\begin{figure}
\centerline{
\includegraphics[width={0.5\textwidth}]{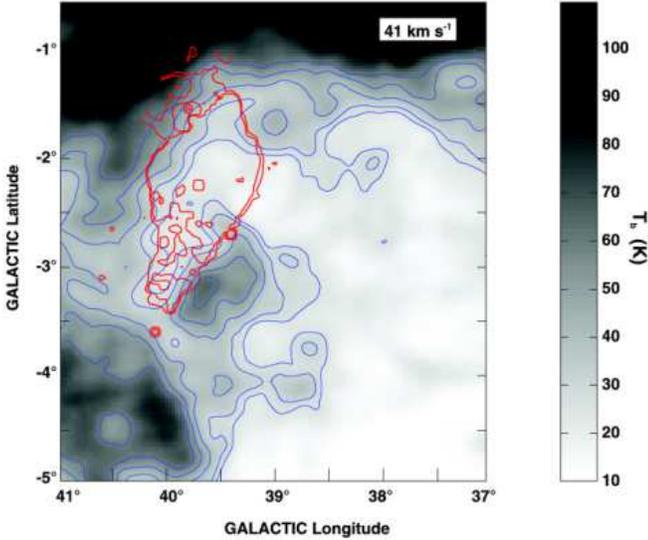}}
\caption{\label{lb_v41}
The GBT 21cm H\,I emission image at 41 km~s$^{-1}$ 
showing the apparent association between H\,I and the W50 
remnant which \citet{Dubner} discovered in lower-resolution data.  
H\,I contours are plotted every 10 K from 30 K to 70 K. 
The location of W50 in the 1.4 GHz radio continuum  
\citep{Dubner} is shown in red.  While we confirm the 
correlation between a void in H\,I emission and the radio 
contours of the eastern section of W50, we believe that it is not 
a physical association, but a coincidence. 
}
\end{figure}

\subsection{+40 km~s$^{-1}$}

\citet{Dubner} detected a void in the H\,I surrounding W50 at a
velocity near 40 km~s$^{-1}$ in maps made at $21\arcmin$
resolution. Figure~\ref{lb_v41} shows the higher angular resolution
GBT image of this velocity range.  The radio continuum coincides with
an H\, I minimum which has a morphology with some superficial
resemblence to that of W50, just as was discovered by \citet{Dubner}
in the lower-resolution data. The inference drawn by Dubner et
al. that the minimum is a void created by the expansion of W50 into
the surrounding medium is, however, quite difficult to understand in
the context of all the other data presented here.

The main difficulty is with the velocity of the void, $V_{LSR} \approx 40$
km~s$^{-1}$. A large
void produced by an SN  should have the velocity of the
ambient ISM into which it expands, in this case $\sim 75$ km~s$^{-1}$.
A void at 40 km~s$^{-1}$  corresponds to a kinematic distance of only 2--3
kpc, whereas all other evidence now places SS\,433 at a 
distance of $5.5$ kpc and a systemic velocity near 75 km~s$^{-1}$.
   Moreover, the H\,I at 40 km~s$^{-1}$ does not seem to be
concentrated into a single cloud, which might, however improbably, be
co-located with SS\,433 yet have a peculiar velocity of --35
km~s$^{-1}$.  Instead, the H\,I seems to lie in a continuous ridge
which is a large-scale interstellar feature, extending
 below the plane to at least $b = -5\arcdeg$.  The
liklihood that the H\,I at 40 km~s$^{-1}$ has a peculiar motion $>30$
km~s$^{-1}$, surrounds SS\,433, and has been evacuated by the SN, seems
low.

We could speculate that the H\,I seen around W50 at 40 km~s$^{-1}$ is
the swept-up shell from the expansion of the SN, accelerated toward us
(and thus from 75 km~s$^{-1}$ to 40 km~s$^{-1}$) by the asymmetric
action of the SS\,433 eastern jet, whose axis is in fact tilted in our
direction, on average, by about $10\arcdeg$ from the plane of the
sky. But there is so much H\,I at 40 km~s$^{-1}$ that it would have
$E_{\rm k} \gg 10^{51}$ ergs in this scenario.  Moreover, the H\,I
ridge at 40 km~s$^{-1}$ does not look like a swept-up wall, but
extends off our image to higher longitude and more negative latitude.
It seems to be a large H\,I cloud and not part of a shell.  On
consideration of energetics and morphology we find it implausible that
the 40 km~s$^{-1}$ feature is associated with SS\,433/W50.

The least gratifying but 
 most likely circumstance given all the evidence is that the apparent 
connection at this velocity 
is a coincidential overlap along our line of sight of an unrelated 
H\,I feature.  Whereas the expanding shell and ring discussed in the 
previous section have both a spatial and kinematic correspondance with
the SS\,433/W50 system, the void at 40 km~s$^{-1}$ matches only in
its general shape.   We conclude that it is unlikely that the 
apparent H\,I void at 40 km~s$^{-1}$ is related to SS\,433.

\subsection{High positive-velocity H\,I}

\citet{KooHeiles} reported the possible detection, though at low
statistical significance, of very weak high-positive-velocity H\,I in
measurements made with an angular resolution of $35\arcmin$ in the
direction of W50. In the \citet{Dubner} model this was assumed to be
material accelerated by W50.  The GBT data show no evidence for an
excess of H\,I toward W50 at velocities $V_{LSR} \gtrsim 120$
km~s$^{-1}$ to a level of $N_{HI} = 0.6 \pm 2.0 \times 10^{18}$
cm$^{-2} \ (1\ \sigma)$, and thus we find no evidence of a
high-velocity receding neutral shell around W50.

\subsection{H\,I at -35 km~s$^{-1}$}

\begin{figure}
\centerline{
\includegraphics[width={0.5\textwidth}]{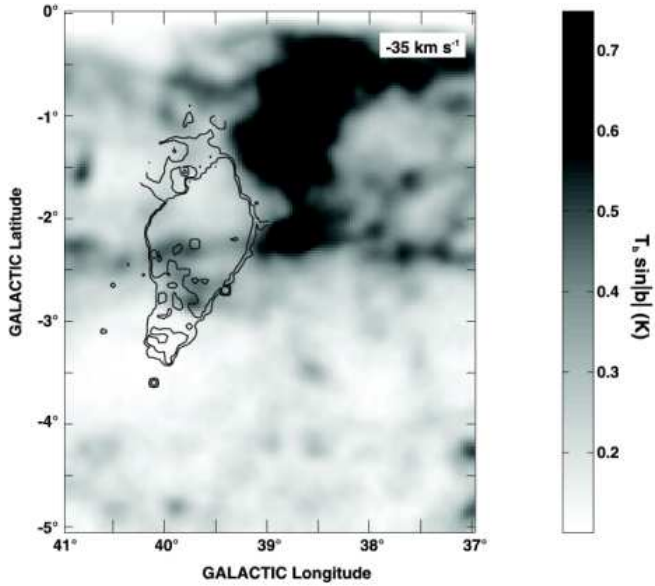}}
\caption{\label{lb_v-35} 
A GBT 21cm H\,I emission image in a 2 km~s$^{-1}$ velocity range 
near --35 km~s$^{-1}$ and the W50 radio contours.  
Here the H\,I intensities have been scaled by $\sin|b|$ to 
reduce the strong brightness gradient with latitude.  We see no 
sign of a concentration of H\,I at this velocity which might be 
associated with material accelerated toward us by the expansion of 
W50. 
}
\end{figure}

\citet{Dubner} suggested that a ridge of H\,I in their data near 
--35 km~s$^{-1}$ might be material accelerated by the shock front
associated with the near-side expansion of W50.  There was also some
evidence from earlier H\,I observations of a shell around W50 at this
velocity \citep{Gosachinskii}.  The GBT image is shown in
Figure~\ref{lb_v-35} with $T_b$ scaled by $\sin|b|$ to reduce the
strong latitude gradient.  We can find no concentration of H\,I that
has a clear association with W50 in this velocity range.

%% figure 15

\subsection{A possible component of the expanding shell 
at --14 km~s$^{-1}$}

\begin{figure}
\centerline{
\includegraphics[width={0.5\textwidth}]{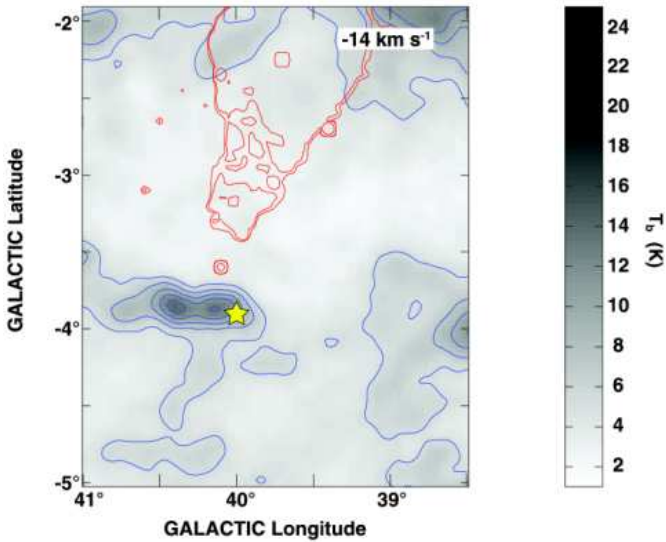}}
%% figure 16
\caption{\label{lb_bubble_v-14}
A GBT H\,I image at $-14$ km~s$^{-1}$ showing a feature which 
lies projected onto the expanding shell; it is labelled `N' in 
Fig.~\ref{vlvb_bubble} and Fig.~\ref{bubblespect}.  The 
H\,I spectrum in Fig.~10 was taken at the location of the yellow star.  
While the spatial coincidence between this feature and the expanding 
shell is strong, there is other emission at the 
same velocity in this field which does not seem related to SS\,433, so 
the overlap is likely a coincidence. 
}
\end{figure}

 Fig.~\ref{lb_bubble_v-14} shows the image of the H\,I emission
 feature labelled `N' in Figs.~9 and 10.  We are doubtful of any
 association with the expanding shell simply because its velocity of
 -14 km~s$^{-1}$ is so extremely different from that of the rest of
 the shell. Nonetheless, it is relatively compact and is projected on
 the centre of the expanding shell, so we consider the consequences of
 its association. This component has a peak $N_{HI} = 8 \times
 10^{19}$ cm$^{-2}$ with a linewidth $\Delta v = 5$ km~s$^{-1}$. The
 position of its peak is $30\fdg34-3\fdg88$ and it extends $30\arcmin$
 in longitude at $<10\arcmin$ in latitude for a size $(50 \times
 \leq16)d_{5.5}$ pc, where $d_{5.5}$ is the distance in units of 5.5
 kpc.  Its H\,I mass is $1.2\pm0.3 \times 10^3 \ d_{5.5}^2 M_{\sun}$,
 and if it has been ejected from the expanding shell, its kinetic
 energy would be $ E_{\rm k} \approx 0.9\pm0.2 \times 10^{50}\
 d_{5.5}^2 $ ergs.

We will not discuss this cloud any further as we consider it unlikely 
to be associated with the expanding shell, whose systemic velocity 
is $71\pm3$  km~s$^{-1}$ and which seems to be a very 
regular structure with a uniform expansion velocity 
of $16$ km~s$^{-1}$. 

\section{The Origin of SS\,433: A Runaway Binary from the Plane}

\begin{figure*}
\centerline{
\includegraphics[width={0.95\textwidth}]{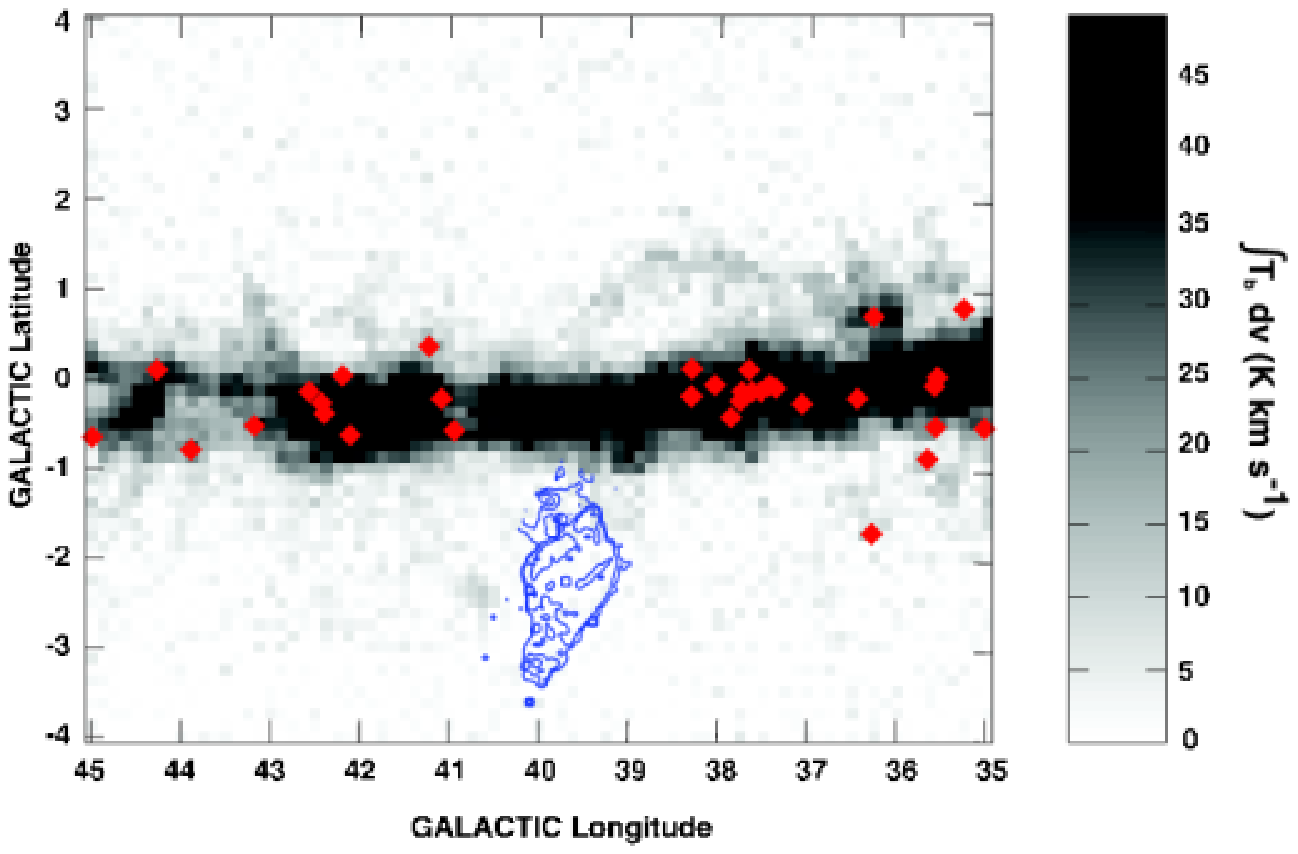}}
\caption{\label{SS433inGalaxy} The 
$^{12}$CO molecular line 
emission integrated over 50-110 km~s$^{-1}$ (gray scale) from 
\citet{Dame}, radio H\,II regions with $V_{LSR}>50$ km s$^{-1}$ 
(red diamonds; \citet{Lockman89, 
LockmanPisanoHoward}),  and the 
radio continuum of SS\,433/W50 \citep{Dubner}, 
showing the relationship between the 
current sites of star formation in the inner Galaxy and the SS\,433 system. 
SS\,433 is located more than 200 pc below the Galactic plane,  well 
away from molecular clouds and other Population I objects.  The absence of 
a molecular cloud  in its vicinity implies that it was formed in 
a cluster in the Galactic plane and then ejected.
}
\end{figure*}

Most studies of the SS\,433 system conclude that it is comprised of a
compact object (possibly a black hole) which likely began its life as
an O star, and a `donor' star of mass $\gtrsim 10 \ M_{\sun}$ (e.g.,
\citet{Hillwig,Fuchsetal}).  By these standards, the system must be
relatively young and belong to Galactic Population I.  It is located, 
however, quite far from the  Galactic plane.  
The young, massive stars in the inner
Galaxy traced by HII regions have a vertical distribution with a
dispersion $\sigma_z = 20$ pc, consistent with the scale height of OB
stars in the solar neighbourhood \citep{LockmanPisanoHoward, Reed,
Maiz}.  At a distance from the Galactic plane 
${\rm z= -215}$ pc, the location of SS\,433 is certainly unusual, 
and as a massive binary it would have attracted notice even if it had
not become an extraordinary system. Its isolation is illustrated in
Figure~\ref{SS433inGalaxy}, which shows the radio continuum with
respect to molecular clouds and HII regions in its area of the
Galaxy. As noted in $\S4$, there is no trace of a molecular cloud at
its longitude, latitude and velocity.  It is thus highly likely that
it was formed near the plane and not at its present location, and 
 must have acquired a peculiar vertical velocity $\gtrsim30$ km~s$^{-1}$ 
to reach z=--215 pc.

%% figure 17

Most, if not all, early-type stars found far from the Galactic plane
are runaways which have been  ejected from their parent cluster 
\citep{Blaauw1961,Gies87}.  Some have a 
 proper motion which allows them to be traced back to 
their origin  (e.g., \citet{Hoogerwerf}). 
Binary stars can also be ejected from young clusters. 
The frequency of binaries among early-type runaway stars is small, 
but not zero \citep{Masonetal,Martin,McSwain}, and theoretical studies 
suggest that binaries containing massive stars can be ejected from 
clusters at velocities $\lesssim 50$  km~s$^{-1}$; 
the more massive the system ejected, the lower the ejection velocity
\citep{LeonardDuncan}.  At the location in the Galaxy of 
SS\,433,  an object launched from ${\rm z=0}$  with a 
vertical velocity $\approx 30$ km~s$^{-1}$   in the Galactic 
potential given by \citet{Wolfire95} will reach the altitude of 
SS\,433 in about 8 Myr.  As importantly, 
because of the long turn-around time, it will stay at $z \approx -200$ pc 
 for  between 8 and 18 Myr after leaving the plane.  This 
corresponds to the main sequence lifetime of stars with 
$M \approx 20 - 25\  M_{\sun}$  \citep{Hirschi}.  
It is thus reasonable that SS\,433 began  in one of the 
young clusters in the Galactic plane, was given a random kick of a few tens 
of km~s$^{-1}$ (comparable to the velocity dispersion of the population 
of runaway OB stars \citep{Stone1991}), 
and made its way as an OB binary to $z \approx -200$ pc where the more massive 
star became a SN  forming W50 and SS\,433.

The other possibility is that the supernova occured while the binary 
was still in the Galactic plane and the SN caused its own ejection
\citep{IbenTutukov}, now with a compact object as one member of the system
\citep{VandenHeuvel}. Arguments for this scenario are summarised by 
\citet{Konigl}, who notes that it would find support if SS\,433 has a
 peculiar velocity directed away from the Galactic plane.

Recent astrometric studies of SS\,433 using the VLBA have been made at
three epochs (A.~Mioduszewski and M.~Rupen, private communication) and
a proper motion has been detected.  Those results will be described in
full elsewhere; here we use them in the discussion of the origin of
SS\,433 and its interaction with the ISM.  The Mioduszewski and Rupen
measurements in March 1998 and August 2005 combine to give a proper
motion $\mu_{\alpha} = -3.5$ and $\mu_{\delta} = -4.6$ milli-arcsec
per year. At a distance of 5.5 kpc and a heliocentric velocity of
$+65$ km~s$^{-1}$ \citep{Hillwig}, we derive velocities relative to
the LSR (standard solar motion) of $V_R = -156$, $V_{\theta} = -57,
V_z = +30$ km~s$^{-1}$, where $R$ is positive radially outward from
the Galactic centre, $V_{\theta}$ is positive in the direction of
Galactic rotation, and $z$ is positive toward the North Galactic Pole.
For a flat rotation curve at $V_{\theta} = 220$ km~s$^{-1}$, the
velocities in a frame corotating at SS\,433 are V(SS\,433) = -17, +225,
+30 km s$^{-1}$ in the R, $\theta$ and z directions, respectively.
SS\,433 thus has peculiar velocities relative to its local standard of
rest of -17, +5, +30 km~s$^{-1}$, with uncertainties of perhaps 50\%
in each value.  The measurements indicate that the total non-circular
motion of SS\,433 is small, $\sim 35$ km~s$^{-1}$, and directed mainly
toward the Galactic plane nearly along the axis of the Western jet.

These results allow us to constrain the history of the system.  If the
O star binary was created in the Galactic plane and ejected before the
SN stage, its age could be $\leq 10$ Myr, where the inequality applies
if the ejection velocity was more than the minimum needed to achieve
its current distance from the plane.  Its current space velocity
presumably reflects the kick that the system got during the SN, and
contains no information on its history prior to the SN.
Alternatively, if the supernova occured in the Galactic plane, and the system
was then ejected, its current velocity reflects its dynamical history.  
It must have reached its greatest distance from the plane, turned
around, and now be returning  at $ +30$ km~s$^{-1}$, which 
 requires a total time of 22 Myr in the 
Galactic potential of \citet{Wolfire95}.

Another factor is that SS\,433 lies at almost the exact centre of W50
measured transverse to the axis of the jets, but is offset by about 4
pc ($5\arcmin$) to the west along the jet axis as measured from the
peak of the X-ray emission \citep{Watson1983}.  Measured from the
maximum extent of the radio lobes it is offset in the same direction
by about 12 pc.  Some asymmetry in the extent of the W50 lobes is expected 
because they are moving nearly perpendicular to the Galactic vertical 
density gradient, though in this case there are suggestions that local 
conditions may be atypical $(\S8.2)$.   The central part of W50 is, 
however, quite circular.  
If we ascribe the offset of SS\,433 to its proper motion, 
 it implies a time from the formation of W50 of
$1 \times 10^5$ y.  This is identical to the lifetime of SS\,433
estimated by \citet{Begelman} adjusting their values for a distance of
5.5 kpc, a jet luminosity of $10^{39}$ ergs~s$^{-1}$ and an ambient
density of 0.5 cm$^{-3}$.

We believe  it is most likely that SS\,433 was formed not far from
where we observe it today, when the more massive member of a 
binary which had been ejected from the plane 
 underwent its SN phase about $10^5$ years ago.

\section{The impact of SS\,433 on the Surrounding Interstellar Medium}

The radio continuum lobes aligned with the SS\,433 jets  evidence
 interaction with the local ISM: measured from SS\,433, the 
lobe directed toward the Galactic plane is  0.7 times the length of
the lobe directed away from the plane.  This is the expected ratio
 if the jet length is limited by swept-up interstellar matter
in an H\,I layer with a vertical density profile given by
\citet{DickeyLockman}.  In the data presented here the most convincing
imprint of this system on the ISM is found at velocities near the
systemic velocity of 75 km~s$^{-1}$: the two H\,I shells, one
expanding and the other static.  While these have some resemblance to
other interstellar shells, in detail they seem to be quite different.

\begin{table}
\centering
\begin{tabular}{lcc}
\hline
& & \\
Property & Expanding & Static \\
& & \\
\hline
$\ell,b$  & $40\fdg2$,--3$\fdg9$ & $39\fdg6$,--1$\fdg8$ \\
z (pc) & --375 & --170  \\
$V_0$ (km~s$^{-1}$)   &  $71\pm3$ & $75\pm3$ \\
$V_{ex}$ (km~s$^{-1}$)  & $16\pm3$ & --- \\
$\Delta V$ (km~s$^{-1}$)  & 5--8  & $\sim 16$ \\
radius (pc)  & $40\pm7$ & $70\pm20$  \\
$\Delta {\rm r}$ (FWHM) (pc)  & $20\pm4$  & $40\pm12$  \\
M$_{\rm H\,I}$ (M$_{\sun})$   & $8\pm3 \times 10^3$ & $3.5 - 10 \times 10^4$ \\
E$_{\rm k}$ (ergs)  & $3\pm1.5 \times 10^{49} $ & --- \\
n$_0$ (cm$^{-3}$) & 0.4 -- 1.2 & 1 -- 2  \\
\hline
\end{tabular}
\caption{Properties of the SS\,433/W50 H\,I Shells calculated for a
  distance of $5.5\pm1$ kpc. 
Note: the property labelled radius measures the distance from the
centre to the peak of emission.} 
\end{table}

\subsection{Origin of the Expanding Shell}

Is the expanding H\,I shell found just off the tip of the Eastern jet
related to, and possibly powered by SS\,433?  The key properties of
this system are summarised in Table~1.  If we were to interpret this
object as a bubble driven by a stellar wind \citep{Castoretal}, for
the observed size and mean density its age would be $\sim 2 \times
10^6$ years and the required mechanical wind luminosity $L_w \sim 2
\times 10^{36}$ ergs s$^{-1}$ or a total power of $10^{50}$ ergs.
This might be supplied by a massive star with a wind velocity of 2000
km~s$^{-1}$ and a mass loss rate of $10^{-6}$ M$_{\sun}$ y$^{-1}$
(e.g., \citet{Kudritzki, Krticka}) but we can find no evidence for
such a star or any other energetic object within the shell, there is
no radio continuum associated with the expanding shell 
in the maps of \citet{Dubner},
and the shell is not visible in the X-ray maps from the ROSAT survey.
Moreover, its location nearly 400 pc below the Galactic plane places
it many scale-heights away from most Population I objects.

 The edge of the shell, however, lies 
exactly on the termination of the radio continuum 
powered by the Eastern jet of SS\,433, its velocity is
close to that of the SS\,433 system, and the projected jet axis goes
precisely through the centre of the H\,I feature.  The jets of SS\,433
have a kinetic luminosity $\sim10^{39}$ ergs s$^{-1}$
\citep{Margon84,PanferovFabrika}, which could supply the necessary
energy in $\sim10^4$ years at an efficiency of $10\%$.  The creation of 
the shell might have been a brief episode in the life of SS\,433, marked 
now only by the fossil record left in  H\,I.  This  
sequence of events is speculative, but if SS\,433 is the 
source of the shell,  the shell age must be  $\sim 10^5$ y,  
 about an order of magnitude less than 
the age derived from simple application of the theory of wind-blown 
bubbles \citep{Castoretal} for an object with its properties.

\subsection{The Static Ring}

The static ring is problematic. It is blended with unrelated
gas, is not symmetric about SS\,433 
but is centred on the upper part of W50, and it
corresponds to the shape of the radio continuum only along the edge of
the western lobe.  Indeed, its basic topology is uncertain.  Higher
resolution H\,I observations are necessary.  There are H\,I shells
observed in the LMC with a size similar to that of the ring, shells
with a low expansion velocity as well, but these usually surround OB
associations, and may not be relevant to the shell around W50.

Simple application of the equations for wind-blown bubbles
\citep{Castoretal} assuming an expansion velocity of $\leq 5$
km~s$^{-1}$, gives the rather large age of $\geq 10^7$ y.  This is
comparable to the free-fall time of an object from the location of
SS\,433 back to the Galactic plane, and is unlikely to be an accurate
description of the age of the ring.  The theory of bubbles and
superbubbles has been reasonably successful when applied to the
interstellar structures formed by winds and SN (e.g.,
\citet{Oey2007}), but it does not seem to give reasonable results for
the static H\,I ring associated with SS\,433/W50.

\subsection{The Average Interstellar Density}

The average H\,I density, $n_0$, is estimated to be 1 -- 2 cm$^{-3}$
at the location of the ring around W50 (${\rm z = -170}$ pc), and 
0.4 -- 1.2 cm$^{-3}$ just below the end of the Eastern jet at 
${\rm z = -374}$ pc. These values are more than an order of magnitude
greater than the densities expected at these 
distances from the Galactic plane
\citep{DickeyLockman}.  It is interesting that there are other
examples of well-defined H\,I shells which also appear to be expanding
into substantially overdense regions
\citep{Stil2004,Gaensler2005}.  Perhaps the prominence of W50 
in the radio continuum is enhanced because of its accidental location 
in a large interstellar H\,I cloud.

\section{Summary Comments}

The new H\,I absorption data reconcile the kinematic distance to
SS\,433 with the distance derived from an analysis of light-travel time 
effects \citep{BlundellBowler}.  
Absorption  at +75 km~s$^{-1}$ is detected in two
separate measurements at the $>10\sigma$ significance level.
 All data now place SS\,433 at a distance 5.5 kpc from the Sun with a
systemic velocity $V_{LSR} \approx 75$ km~s$^{-1}$. The HII
region S74 and the molecular cloud at 30 km~s$^{-1}$, both of which overlap    
parts of the W50 remnant, have no association with the system and 
most likely lie in the foreground, as suspected by
\citet{Margon84}. We also find no convincing evidence for an association of
any H\,I with W50 at velocities other than near 75 km~s$^{-1}$.

The determination of a modest proper motion of 35 km~s$^{-1}$ 
directed nearly along the Western jet (i.e. toward the Galactic 
plane) suggests that  SS\,433  originated 
as an O-star binary which was ejected early in its life from a 
cluster near the Galactic plane with a velocity away from the Galactic 
plane $\gtrsim 30$ km~s$^{-1}$. The SN of the more massive star then 
occured close to where we now observe the system, and in that event it 
was given a modest kick back toward the Galactic plane.  
The location of SS\,433 near the centre of symmetry 
of the radio continuum source W50 suggests that the 
 SN  occured  $10^5$ years ago.

There is no molecular cloud (as traced by CO emission) associated with
the system, but there seem to be two H\,I features: a static ring or
shell around the upper part of W50, and an expanding shell touching
the end of the Eastern lobe.  There are many H\,I shells in the ISM
\citep{Menon,Heiles79,McClure2002, Ehlerova} often seen surrounding
H\,II regions or clusters of young stars, in other instances without
an identifiable source of energy \citep{Stil2004}, but well-developed
H\,I shells around supernova remnants are rare.  Studies of individual
remnants often show patchy fragmented structures, perhaps because
remnants expand into a complex ISM \citep{Koo93,Kooetal2004,Uyaniker}.
It may be significant that the average interstellar density derived
for both H\,I features around W50 is about an order of magnitude
greater than what would be expected at their distance from the
Galactic plane.  This seems to be a trend for some H\,I shells which
are not near regions of star formation
\citep{Gaensler2005,Stil2004}.

The expanding shell is quite interesting, for its measured kinetic
energy is so large ($3 \times 10^{49}$ ergs) that it must have been
formed by some energetic event, but it is quite far from the Galactic
plane and there is no nearby candidate energy source except for
SS\,433. It is possible that the shell is a fossil remnant of a time
when the Eastern jet extended out past the current radio continuum
boundries -- the jet has enough energy to create a bubble of this sort
in  $\sim 10^4$ years.  On the other hand, there are examples of H\,I
shells with unidentified energy sources, and it may be a challenge to
understand how the kinetic energy of a collimated jet could produce a
relatively symmetric expanding shell.

The SS\,433 system is ideally situated for the study of the
interaction between  
energetic events and the interstellar medium: it is separated from 
most confusing sources, accessible at many wavelengths, and has 
resonably well understood energetics.  
As noted by \citet{Konigl}, the SS\,433 jets may offer a unique probe of the 
ISM.  Further study of H\,I in this region should be rewarding.

\section*{Acknowledgments}

The National Radio Astronomy Observatory is a facility of the National
Science Foundation operated under cooperative agreement with
Associated Universities, Inc.  We thank Alison Peck for assistance
with the VLA observations, Mark Heyer \& Chris Brunt for obtaining the
FCRAO $^{12}$CO data and sharing it with us, and Amy Mioduszewski \&
Michael Rupen for sharing their proper motion data.  We also thank
Vivek Dwahan, Bob Benjamin, and Mike Shull for useful discussions.

% the figures

\bsp
\label{lastpage}


\begin{thebibliography}{}

\bibitem[Ma{\'{\i}}z-Apell{\'a}niz(2001)]{Maiz} 
Ma{\'{\i}}z-Apell{\'a}niz, J.\ 2001, \aj, 121, 2737 
%% The Spatial Distribution of O-B5 Stars in the Solar 
%%               Neighbourhood as Measured by Hipparcos

\bibitem[Begelman et al.(1980)]{Begelman} Begelman, M.~C., 
Hatchett, S.~P., McKee, C.~F., Sarazin, C.~L., \& Arons, J.\ 1980, \apj, 
238, 722 
%% Beam models for SS 433

\bibitem[Blaauw(1961)]{Blaauw1961} Blaauw, A.\ 1961, \bain, 15, 265
%% On the origin of the O- and B-type stars with high velocities 
%%   (the "run-away" stars), and some related problems

\bibitem[Blundell \& Bowler (2004)]{BlundellBowler} Blundell, K.M., \& 
Bowler, M.G. 2004, \apj, 616, L159
%% Symmetry in the changing jets of SS\,433 and its true distance from us

\bibitem[Brand \& Blitz (1993)]{BrandBlitz} Brand, J., \& Blitz, L. 1993, 
A\&A, 275, 67
%% The velocity field of the outer Galaxy


\bibitem[Brinkmann et al.(2006)]{Brinkmann2006} Brinkmann, W., Pratt, 
G.~W., Rohr, S., Kawai, N., \& Burwitz, V.\ 2006, ArXiv Astrophysics 
e-prints, arXiv:astro-ph/0610781 
%% XMM-Newton observations of the eastern jet of SS433

\bibitem[Burton (1992)]{Burtonrot} Burton, W.B. 1992, 
in The Galactic Interstellar Medium, Saas-Fee Advanced Course 21, ed. D.
Pfenniger, \& P. Bartholdi, (Springer-Verlag), p. 1

\bibitem[Castor et al.(1975)]{Castoretal} Castor, J., McCray, R., 
\& Weaver, R.\ 1975, \apjl, 200, L107 

\bibitem[Clemens (1985)]{Clemens} Clemens, D.P. 1985, \apj, 295, 422
%% Mass-Stony Brook Galactic Plane CO survey: the galactic disc rotation curve

\bibitem[Cioffi et al.(1988)]{Cioffi} Cioffi, D.~F., McKee, 
C.~F., \& Bertschinger, E.\ 1988, \apj, 334, 252 
%% Dynamics of radiative supernova remnants

\bibitem[Dame et al.(2001)]{Dame} Dame, T.~M., Hartmann, D., 
\& Thaddeus, P.\ 2001, \apj, 547, 792 

\bibitem[Delhaye(1965)]{Delhaye} Delhaye, J.\ 1965, Galactic Structure, 
 ed. A. Blaauw \& M. Schmidt, Univ. Chicago Press, 61 

\bibitem[Dickey et al. (1983)]{Dickey83} Dickey, J. M., Kulkarni, S. R.,
 Heiles, C. E., \& van Gorkom, J. H. 1983, \apjs 53,  591
%% A survey of H I absorption at low latitudes

\bibitem[Dickey \& Lockman(1990)]{DickeyLockman} Dickey, J.~M., 
\& Lockman, F.~J. 1990, \araa, 28, 215

\bibitem[Downes et al.(1986)]{Downes1986} Downes, A.~J.~B., Pauls, 
T., \& Salter, C.~J.\ 1986, \mnras, 218, 393
%% The supernova remnant W50 at 5 GHz
 
\bibitem[Dwarkadas(2006)]{Dwarkadas2006} Dwarkadas, V.~V.\ 2006, \apss, 537
%% Hydrodynamics of Supernova Evolution in the Winds of Massive Stars

\bibitem[Dubner et al. (1998)]{Dubner} Dubner, G.M., Holdaway, M., 
Goss, W.M., \& Mirabel, I.F. 1998, \aj, 116, 1842 
%% A high-resolution radio study of the W50-SS\,433 system and the 
%% surrounding medium

\bibitem[Ehlerov{\'a} \& Palou{\v s}(2005)]{Ehlerova} 
Ehlerov{\'a}, S., \& Palou{\v s}, J.\ 2005, \aap, 437, 101
%% H I shells in the outer Milky Way
 
\bibitem[Englmaier \& Gerhard (2006)]{EnglmaierGerhard06} 
Englmaier, P., \& Gerhard, O. 2006, 
Celest. Mech. and Dynam. Astr.,  94, 369
%%   Milky Way Gas Dynamics

\bibitem[Forbes (1989)]{Forbes} Forbes, D. 1989,  A\&AS, 77, 439
%% Photometry and Spectroscopy of stars in northern HII regions

\bibitem[Fuchs et al.(2006)]{Fuchsetal} Fuchs, Y., Koch Miramond, 
L., \& {\'A}brah{\'a}m, P.\ 2006, \aap, 445, 1041
%% SS 433: a phenomenon imitating a Wolf-Rayet star

\bibitem[Gaensler et al.(2005)]{Gaensler2005} Gaensler, B.~M., 
McClure-Griffiths, N.~M., Oey, M.~S., Haverkorn, M., Dickey, J.~M., \& 
Green, A.~J.\ 2005, \apjl, 620, L95 
%% A Stellar Wind Bubble Coincident with the Anomalous X-Ray Pulsar 
%  1E 1048.1-5937:  ...

\bibitem[Gies(1987)]{Gies87} Gies, D.~R.\ 1987, \apjs, 64, 545 

\bibitem[Gosachinskii \& Khersonskii(1987)]{Gosachinskii}Gosachinskii, 
I.~V., \& Khersonskii, V.~K.\ 1987, Soviet Astronomy, 31, 621 
%% Distribution of Neutral Hydrogen in the Vicinity of the Supernova 
%%    Remnant 3C396

\bibitem[Heiles(1979)]{Heiles79} Heiles, C.\ 1979, \apj, 229, 533 
%% HI Shells and Supershells

\bibitem[Heyer, Williams, \& Brunt (2006)]{Heyer} Heyer, M.H., 
Williams, J.P., \&   Brunt, C.M. 2006, \apj, 643, 956
%% Turbulent Gas flows in the Rosette and G216-2.5 mol clouds

\bibitem[Hillwig et al.(2004)]{Hillwig} Hillwig, T.~C., Gies, 
D.~R., Huang, W., McSwain, M.~V., Stark, M.~A., van der Meer, A., \& Kaper, 
L.\ 2004, \apj, 615, 422 
%% Identification of the Mass Donor Star's Spectrum in SS 433

\bibitem[Hirschi et al.(2004)]{Hirschi} Hirschi, R., Meynet, 
G., \& Maeder, A.\ 2004, \aap, 425, 649 
%%  Stellar evolution with rotation. XII. Pre-supernova models

\bibitem[Hjellming \& Johnston (1981a)]{Hjellming81} Hjellming, R.M., 
\& Johnston, K.J. 1981a, \nat, 290, 100
%% Structure, strength, and polarization changes in radio source SS\,433

\bibitem[Hjellming \& Johnston(1981b)]{Hjellming81a} Hjellming, 
R.~M., \& Johnston, K.~J.\ 1981b, \apjl, 246, L141 
%% An analysis of the proper motions of SS 433 radio jets

\bibitem[Hoogerwerf et al.(2001)]{Hoogerwerf} Hoogerwerf, R., de 
Bruijne, J.~H.~J., \& de Zeeuw, P.~T.\ 2001, \aap, 365, 49 
%% On the origin of the O and B-type stars with high velocities. 
%%  II. Runaway stars and pulsars ejected from the nearby young stellar groups

\bibitem[Huang, Dame \& Thaddeus (1983)]{Huang} Huang, Y.-L., Dame, T.M., \& 
Thaddeus, P. 1983, \apj, 272, 609
%% A Large Molecular Cloud toward the SNR W50 and SS\,433

\bibitem[Iben \& Tutukov(1997)]{IbenTutukov} Iben, I.~J., \& 
Tutukov, A.~V.\ 1997, \apj, 491, 303 
%% On Space Velocities of Binary Stars in Which One Component 
%%	Has Experienced a Supernova Explosion

\bibitem[Kim et al.(1999)]{Kimetal} Kim, S., Dopita, M.~A., 
Staveley-Smith, L., \& Bessell, M.~S.\ 1999, \aj, 118, 2797 
%% HI shells in the large magellanic cloud

\bibitem[K\"{o}nigl(1983)]{Konigl} K\"{o}nigl, A.\ 1983, \mnras, 205, 471 
%% W50 - A stellar-wind bubble in a three-phase interstellar medium?

\bibitem[Koo \& Heiles (1991)]{KooHeiles} Koo, B.-C., \& Heiles, C. 1991, 
\apj, 382, 204
%% A survey of HI 21 cm emission lines toward supernova remnants

\bibitem[Koo et al.(1993)]{Koo93} Koo, B.-C., Yun, M.-S., Ho, 
P.~T.~P., \& Lee, Y.\ 1993, \apj, 417, 196 
%% Interaction between the Supernova Remnant CTB 80 and the Ambient 
%%   Interstellar Medium: H I and CO Observations

\bibitem[Koo \& Kang(2004)]{2004MNRAS.349..983K} Koo, B.-C., \& Kang, 
J.-h.\ 2004, \mnras, 349, 983
%% Visibility of old supernova remnants in HI 21-cm emission line

\bibitem[Koo et al.(2004)]{Kooetal2004} Koo, B.-C., Kang, J.-H., 
\& McClure-Griffiths, N.~M.\ 2004, Journal of Korean Astronomical 
Society, 37, 61 
%% HI 21 cm Emission Line Study of Southern Galactic Supernova Remnants

\bibitem[Krticka \& Kubat(2007)]{Krticka} Krticka, J., \& 
Kubat, J.\ 2007, ArXiv Astrophysics e-prints, arXiv:astro-ph/0701411 

\bibitem[Kudritzki \& Puls(2000)]{Kudritzki} Kudritzki, R.-P., \& 
Puls, J.\ 2000, \araa, 38, 613 

\bibitem[Leonard \& Duncan(1990)]{LeonardDuncan} Leonard, P.~J.~T., 
\& Duncan, M.~J.\ 1990, \aj, 99, 608 
%% Runaway stars from young star clusters containing initial binaries. 
%%  II - A mass spectrum and a binary energy spectrum

\bibitem[Lockman(1989)]{Lockman89} Lockman, F.~J.\ 1989, \apjs, 71, 469 
%% A survey of HII regions in the inner Galaxy

\bibitem[Lockman \& Condon (2005)]{LockmanCondon} Lockman, F.J., 
\& Condon, J.J. 2005, \aj, 129, 1968
%% The Spitzer space telescope first look survey: neutral hydrogen emission

\bibitem[Lockman, Pisano, \& Howard (1996)]{LockmanPisanoHoward} 
Lockman, F.J., Pisano, D.J., \& Howard, G.J. 1996, \apj, 472, 173
%% Detection of 130 ``diffuse'' galactic HII regions 

\bibitem[Margon (1984)]{Margon84} Margon, B. 1984, \araa, 22, 507
%% Observations of SS\,433

\bibitem[Margon et al. (1979)]{Margon79} Margon, B., Ford, H.C., Katz, 
J.I., Kwitter, K.B., Ulrich, R.K., Stone, R.P.S., \& Klemola, A. 1979, 
\apj, 230, L41
%% The Bizarre spectrum of SS\,433

\bibitem[Martin(2006)]{Martin} Martin, J.~C.\ 2006, \aj, 131, 3047 
%% The Origins and Evolutionary Status of B Stars Found Far from the 
%%	Galactic Plane. II. Kinematics and Full Sample Analysis

\bibitem[Mason et al.(1998)]{Masonetal} Mason, B.~D., Gies, 
D.~R., Hartkopf, W.~I., Bagnuolo, W.~G., Jr., ten Brummelaar, T., \& 
McAlister, H.~A.\ 1998, \aj, 115, 821
%%  ICCD speckle observations of binary stars. 
%%     XIX - an astrometric/spectroscopic survey of O stars

\bibitem[McClure-Griffiths et al.(2002)]{McClure2002} 
McClure-Griffiths, N.~M., Dickey, J.~M., Gaensler, B.~M., 
\& Green, A.~J.\ 2002, \apj, 578, 176 
%% The Galactic Distribution of Large H I Shells

\bibitem[McSwain et al.(2007)]{McSwain} McSwain, M.~V., 
Boyajian, T.~S., Grundstrom, E.~D., \& Gies, D.~R.\ 2007, \apj, 655, 473 
%% A Spectroscopic Study of Field and Runaway OB Stars

\bibitem[Menon(1958)]{Menon} Menon, T.~K.\ 1958, \apj, 127, 28 
%% Interstellar Structure of the Orion Region. I.

\bibitem[Murdin, Clark, \& Martin (1980)]{MurdinClarkMartin}
Murdin, P., Clark, D.H., \& Martin, P.G. 1980, \mnras, 193, 135
%% The optical spectrum of SS\,433

\bibitem[Oey(2004)]{Oey2004} Oey, M.~S.\ 2004, \apss, 289, 269
%% Superbubble Activity in Star-Forming Galaxies
 
\bibitem[Oey(2007)]{Oey2007} Oey, M.~S.\ 2007, IAU Symposium, 237, 106 
%% Towards resolving the evolution of multi-supernova superbubbles

\bibitem[Panferov \& Fabrika(1997)]{PanferovFabrika} Panferov, A.~A., 
\& Fabrika, S.~N.\ 1997, Astronomy Reports, 41, 506 

\bibitem[Reich et al. (1990)]{Reich} Reich, W., F\"urst, E., Reich, P., 
\& Reif, K. 1990, A\&AS, 85,  633

\bibitem[Reed(2000)]{Reed} Reed, B.~C.\ 2000, \aj, 120, 314 
%% New estimages of the scale height and surface density of OB stars
%%  in the solar neighbourhood

\bibitem[Stil et al.(2004)]{Stil2004} Stil, J.~M., Taylor, 
A.~R., Martin, P.~G., Rothwell, T.~A., Dickey, J.~M., \& McClure-Griffiths, 
N.~M.\ 2004, \apj, 608, 297 
%% GSH 23.0-0.7+117: A Neutral Hydrogen Shell in the Inner Galaxy

\bibitem[Stirling et al.(2002)]{Stirling} Stirling, A.~M., 
Jowett, F.~H., Spencer, R.~E., Paragi, Z., Ogley, R.~N., \& Cawthorne, 
T.~V.\ 2002, \mnras, 337, 657 
%% Radio-emitting component kinematics in SS433

\bibitem[Stone(1991)]{Stone1991} Stone, R.~C.\ 1991, \aj, 102, 333
%%  The space frequency and origin of the runaway O and B stars

\bibitem[van Gorkom, Goss, \& Shaver (1979)]{vanG79} van Gorkom, J.H., 
Goss, W.M., \& Shaver, P.A. 1979, A\&A, 82, L1
%% HI Absorption in the direction of SS\,433

\bibitem[van den Heuvel et al.(1980)]{VandenHeuvel} van den Heuvel, 
E.~P.~J., Ostriker, J.~P., \& Petterson, J.~A.\ 1980, \aap, 81, L7 
%% An early-type binary model for SS433

\bibitem[van Gorkom et al. (1982)]{vanG82} 
van Gorkom, J.H., Goss, W.M., Seaquist, E.R., \& Gilmore, W.S. 
1982, \mnras, 198, 757
%% HI absorption distances to four point sources near supernova remnants

\bibitem[Vel{\'a}zquez \& Raga(2000)]{Velazquez} Vel{\'a}zquez, 
P.~F., \& Raga, A.~C.\ 2000, \aap, 362, 780 
%% A numerical simulation of the W 50-SS 433 system

\bibitem[Watson et al.(1983)]{Watson1983} Watson, M.~G., 
Willingale, R., Grindlay, J.~E., \& Seward, F.~D.\ 1983, \apj, 273, 688 
%%The X-ray lobes of SS 433

\bibitem[Weaver et al.(1977)]{Weaveretal} Weaver, R., McCray, R., 
Castor, J., Shapiro, P., \& Moore, R.\ 1977, \apj, 218, 377 
%% Interstellar Bubbles II. Structure and Evolution

\bibitem[Weiner \& Sellwood (1999)]{WeinerSellwood} Weiner, B.J., \& 
Sellwood, J.A. 1999, \apj,  524, 112
%% The Properties of the Galactic Bar Implied by Gas Kinematics 
%%      in the Inner Milky Way

\bibitem[Wolfire et al.(1995)]{Wolfire95} Wolfire, M.~G., McKee, 
C.~F., Hollenbach, D., \& Tielens, A.~G.~G.~M.\ 1995, \apj, 453, 673 
%% The Multiphase Structure of the Galactic Halo: 
%%	High-Velocity Clouds in a Hot Corona

\bibitem[Yamamoto et al. (1999)]{Yamamoto} Yamamoto, F., Hasegawa, T., 
Morino, J., Handa, T., Sawada, T., \& Dame, T.M. 1999, in Star
Formation 1999, Proceedings of Star Formation 1999, held in Nagoya,
Japan, June 21 - 25, 1999, Editor: T. Nakamoto, Nobeyama Radio
Observatory, p. 110-111
%% Supernova Remnant-molecular cloud interactions: the cases of HB21, W66 
%% and W50

\bibitem[Yar-Uyaniker et al.(2004)]{Uyaniker} Yar-Uyaniker, A., 
Uyaniker, B., \& Kothes, R.\ 2004, \apj, 616, 247 
%% Distance of Three Supernova Remnants from H I Line Observations in a 
%% Complex Region: G114.3+0.3, G116.5+1.1, and CTB 1 (G116.9+0.2)


\bibitem[Zealey et al.(1980)]{Zealey1980} Zealey, W.~J., Dopita, 
M.~A., \& Malin, D.~F.\ 1980, \mnras, 192, 731 
%% The interaction between the relativistic jets of SS433 
%%   and the interstellar medium	

\end{thebibliography}
\end{document}